\documentclass{emulateapj}

\shorttitle{Models of the Intracluster Medium}
\shortauthors{McCarthy et al.}
\submitted{To appear in the Astrophysical Journal (received 03/05/04,
accepted 06/10/04)}

\begin{document} 

\title{Models of the ICM with Heating and Cooling: Explaining the 
Global and Structural X-ray Properties of Clusters}

\author{Ian G. McCarthy$^{1}$, Michael L. Balogh$^{2,3}$, Arif Babul$^{1}$, 
Gregory B. Poole$^{1}$, and Donald J. Horner$^{4}$}

\affil{$^1$Department of Physics \& Astronomy, University of Victoria, Victoria, BC, 
V8P 1A1, Canada; mccarthy@uvastro.phys.uvic.ca, babul@uvic.ca, 
gbpoole@uvastro.phys.uvic.ca}
\affil{$^2$Department of Physics, University of Durham, Durham, DH1 3LE, UK; 
m.l.balogh@durham.ac.uk}
\affil{$^3$Present address: Department of Physics, University of Waterloo,
Waterloo, ON, N2L 3G1, Canada}
\affil{$^4$Department of Astronomy, University of Massachusetts, Amherst, MA 
01003; horner@astro.umass.edu}

\begin{abstract}
Non-radiative simulations which only include heating due to gravitational 
processes fail to match the observed mean X-ray properties of galaxy 
clusters.  As a result, there has recently been increased 
interest in models in which either radiative cooling or entropy injection 
(and/or redistribution) play a central role in mediating the thermal and 
spatial properties of the intracluster medium.  Both sets of models can 
account for the mean global properties of clusters.  Radiative cooling alone, 
however, results in fractions of cold/cooled baryons in excess of 
observationally established limits.  On the other hand, the simplest 
entropy injection models, by design, do not treat the ``cooling core'' 
structure present in many clusters and cannot account for declining entropy 
profiles towards cluster centers revealed by recent high resolution X-ray 
observations.  We consider models that marry radiative cooling with entropy 
injection, and confront model predictions for the global and structural 
properties of massive clusters with the latest X-ray data.  The models 
successfully and simultaneously reproduce the observed $L-T$ and $L-M$ 
relations, yield detailed entropy, surface brightness, and temperature profiles 
in excellent agreement with observations, and predict a cooled gas fraction 
that is consistent with observational constraints.  More interestingly, the 
model provides a possible explanation for the significant intrinsic scatter 
present in the $L-T$ and $L-M$ relations.  The model also offers a natural way 
of distinguishing between clusters classically identified as ``cooling flow'' 
clusters and the relaxed ``non-cooling flow'' clusters.   The former correspond 
to systems that 
experienced only mild levels ($\lesssim 300$ keV cm$^2$) of entropy injection, 
while the latter are identified as systems that had much higher entropy 
injection.  The dividing line in entropy injection between the two categories 
corresponds roughly to the cooling threshold for massive clusters.  This 
finding suggests that entropy injection may be an important, if not the 
primary, factor in determining the class a particular cluster will belong 
to.  These results also suggest that the previously identified relationship 
between inferred cooling flow strength and the dispersion in the $L-T$ relation 
is a manifestation of the distribution of cluster entropy injection levels.  
This is borne out by the entropy profiles derived from {\it Chandra} and {\it 
XMM-Newton}.  Finally, the model predicts a relationship between a 
cluster's central entropy and its core radius, the existence of which we 
confirm in the observational data.
\end{abstract}


\section{Introduction}

It has been made increasingly apparent in recent years that theoretical models 
of cluster formation and evolution that incorporate gravitationally-driven 
processes alone fail to match the observed global X-ray properties of clusters 
(e.g., Kaiser 1991, Tozzi \& Norman 2001; Babul et al. 2002).  More recently, 
McCarthy et al. (2003a, 2003b) showed that such models are also incompatible 
with the observed Sunyaev-Zeldovich (SZ) effect properties of distant clusters.  
This discord between theory and observations has motivated a number of authors 
to examine the potential role of {\it non}-gravitational gas physics, usually 
in the form of radiative cooling, entropy injection$^5$ (e.g., from AGN or 
galactic winds), or both cooling and entropy injection (e.g., Kaiser 1991; 
Balogh, Babul, \& Patton 1999; Bryan 2000; Tozzi \& Norman 2001; Borgani et al. 
2001; Voit \& Bryan 2001; Babul et al. 2002; Wu \& Xue 2002; Voit et al. 2002; 
Dav\'{e}, Katz, \& Weinberg 2002; Voit et al. 2003; Oh \& Benson 2003).  These 
models generally compare more favorably to the data [e.g., to the observed 
luminosity-temperature ($L-T$) and luminosity-mass ($L-M$) relations] than the 
standard `non-radiative' model, but it still remains somewhat unclear as to 
which of these models --- those with entropy injection alone, those with 
cooling alone, or those with entropy injection plus cooling --- best reflect 
the true nature of clusters.  

\footnotetext[5]{We also regard mechanisms such as thermal conduction and 
turbulent mixing (e.g., Narayan \& Medvedev 2001; Kim \& Narayan 2003) as 
sources of `entropy injection' since they transfer heat to the cluster center.  
Strictly speaking, however, these processes do not really introduce new entropy 
into the system, they merely {\it redistribute} the cluster's (pre-existing) 
entropy.} 

Part of the reason for the ambiguity undoubtedly arises from the fact that there 
is a relatively large amount of scatter present in the observed X-ray scaling 
relations of clusters, in particular the $L-T$ and $L-M$ relations.  Because of 
the large scatter, the various non-gravitational models, which typically 
predict similar mean global properties, are essentially indistinguishable (see, 
e.g., Voit et al. 2002; Balogh et al. 2004).

The origin of the scatter in the observed relations is uncertain and has 
received very little attention from a theoretical modeling point of view.  It 
is clear that at least some of the dispersion is due to real physical 
differences in the properties of clusters of a given mass (as the observed 
scatter cannot be explained by measurement uncertainty) and, therefore, 
any realistic theoretical model of cluster evolution that seeks to make 
precise predictions must account for it.  Understanding the 
origin of the scatter is of considerable importance to studies seeking to use 
clusters for precision cosmological tests, such as the determination of 
the matter power spectrum normalization, $\sigma_8$ (see, e.g., Smith et al. 
2003; Balogh et al. 2004).

An examination of whether or not the intrinsic scatter can be accounted for by 
non-gravitational 
gas physics, therefore, is one of the primary goals of the 
present paper.  We demonstrate below that {\it the scatter in the $L-T$ 
and $L-M$ relations is inconsistent with entropy injection only or cooling only 
models}.  However, the scatter can be 
accounted for by a model that includes both entropy injection and radiative 
cooling.  Moreover, our analysis indicates variations in the efficiency of 
entropy injection across the cluster population.  Combined cooling + entropy 
injection models also have the advantage of not being subject to the 
``overcooling'' problems that plague the purely radiative cooling models.

While detailed studies of the global properties of clusters have taught us much 
(and continue to teach us) about the intracluster medium (ICM) and clusters in 
general, a potentially much more powerful test is comparisons between observed 
and predicted {\it structural} properties, such as entropy, temperature, and 
surface brightness profiles.  The influx of new high spatial and high 
spectral resolution X-ray data from {\it Chandra} and {\it XMM-Newton} now 
affords us the opportunity to make such comparisons.  A second goal of this 
study, therefore, is to confront theoretical models that include entropy 
injection and/or cooling with new high resolution data.  We note that early 
results from {\it Chandra} and {\it XMM-Newton} show no signs of the large 
isentropic cores in groups and clusters predicted by generic injection only 
models (e.g., David et al. 2001; Pratt \& Arnaud 2003; Mushotzky et al. 2003).  
This implies that other processes, possibly radiative cooling, are also 
important, at least for some clusters.  Indeed, we demonstrate that models with 
both radiative cooling and entropy injection are able to match the observed 
structural properties of massive clusters and simultaneously account for the 
$L-T$ and $L-M$ relations.  Interestingly, the theoretical systems with only 
mild levels of entropy injection look remarkably like ``cooling flow'' (CF) 
clusters$^6$, whereas systems with high levels exhibit the typical 
characteristics of ``non-cooling flow'' (NCF) clusters.     

\footnotetext[6]{The designation {\it ``cooling flow'' cluster} refers to a 
system that has a sharply rising surface brightness profile and, normally, a 
declining temperature profile towards the center.  These observational 
characteristics have typically been interpreted as manifestations of an ICM 
that is radiatively cooling on short timescales.  The cooling gas flows inward 
toward the cluster center (hence, the name cooling flow).  When we use the 
phrase ``cooling flow'' (in quotation marks) we are referring to the 
observational characteristics and not a physical model.}

The present paper is organized as follows.  In \S 2, we extend the models of 
Babul et al. (2002) to include a realistic treatment of radiative cooling.  A 
general discussion of how radiative cooling modifies the properties of the 
models is given in \S 3.  In \S 4, the physical origin of the scatter in the 
$L-T$ and $L-M$ relations is explored.  Comparisons of structural properties 
between the various theoretical models and high quality {\it Chandra} and {\it 
XMM-Newton} data are made in \S 5.  Finally, in \S 6 and \S7, we summarize and 
discuss our results.

The models considered below were developed in a flat $\Lambda$CDM cosmology 
with $h=0.75$, $\Omega_m = 0.3$ and $\Omega_b = 0.020 h^{-2}$, which is a close 
match to current estimates, including those from {\it WMAP} (Spergel et al. 
2003).

\section{Cluster models with radiative cooling}

The primary goal of this paper is to explore how entropy injection and radiative 
cooling influence the evolution of the ICM and to confront these models with 
new high quality X-ray data.  We have already performed a thorough analysis 
of how entropy injection alone modifies the ICM (e.g., Balogh, Babul, \& Patton 
1999; Babul et al. 2002; McCarthy et al. 2002a, 2003a, 2003b, 2003c).
This model sought to explain the properties of groups and clusters {\it minus} 
the ``cooling flow'' component, if any, and therefore, explicitly ignored 
radiative cooling.  We refer the reader to Babul et al. (2002), in 
particular, for an in-depth discussion of the model, including an examination 
of the possible sources of the entropy injection.  We re-examine the issue of 
sources of non-gravitational entropy in \S 6 of the present paper.

Since we have examined the effects of entropy injection in detail, the current 
section is devoted to an examination of the effects of radiative cooling.  
First, however, a short discussion of the initial conditions prior to cooling 
is given so that we may gauge how cooling modifies things.

\subsection{Initial conditions}

There have been claims that radiative cooling alone may explain the deviations 
of clusters from self-similarity (e.g., Bryan 2000; Wu \& Xue 2002; Dav\'{e}, 
Katz, \& Weinberg 2002) or, at least, that it plays the dominant role in the 
breaking of self-similarity (e.g., Voit \& Ponman 2003).  In order to explore 
this possibility, we examine the effects of cooling on the simple isothermal 
model of Babul et al. (2002).  As its name implies, this isothermal model 
assumes that initially (i.e., before any cooling) the ICM has a constant 
temperature, which is set to the cluster virial temperature.  The intracluster 
gas is in hydrostatic equilibrium within a gravitationally-dominant dark 
matter halo that has a density profile which matches those found in recent 
high resolution numerical simulations (e.g., Moore et al. 1999; Lewis et al. 
2000).  In order to solve for hydrostatic equilibrium, it is assumed that the 
cluster is a typical region of the universe in terms of the mixture of dark 
matter and baryons, i.e., the ratio of gas mass to total mass within cluster's 
maximum radius, $r_{\rm halo}$, is given by $\Omega_b/\Omega_m$.  In terms of 
global properties, this simple model has been shown to be in excellent 
agreement with self-similar predictions (i.e., it predicts $L \propto T^2$ and 
$L \propto M^{4/3}$) and with the results of non-radiative simulations, such 
as those performed by Evrard, Metzler, \& Navarro (1996).  The simplicity of 
this model makes it particularly suitable for analysis.  We, therefore, adopt 
it as the baseline model to gauge the impact of cooling.

While our simple isothermal model predicts {\it global} properties that are 
very similar to those seen in non-radiative simulations, it is clear that there 
are some differences between the two in terms of predicted {\it structural} 
properties.  For example, Lewis et al. (2000) and, more recently, Loken et al. 
(2002) find non-isothermal temperature profiles in their non-radiative 
(``adiabatic'') simulations, with the gas temperature dropping by more than a 
factor of 2 from the cluster center to its periphery.  Thus, it is reasonable 
to ask whether or not the isothermal model represents a fair baseline model.  
In order to test this, we turn to the study of Voit et al. (2003).  Using the 
numerical simulations of Bryan \& Voit (2001) (which were run with the same 
adaptive mesh refinement (AMR) code used by Loken et al. 2002), Voit et al. 
(2003) 
showed that the dimensionless entropy profiles of simulated non-radiative 
clusters are approximately self-similar.  We use their self-similar entropy 
profile, which was kindly provided in electronic form by G. M. Voit and G. L. 
Bryan, together with a realistic dark matter density profile (the same profile 
used in the isothermal model) to construct a second baseline model to which we 
can compare our isothermal model.  As expected, the global properties of the 
isothermal model and the non-radiative Voit \& Bryan clusters are quite 
similar but there are some differences between the predicted temperature and 
density profiles.  However, when we allow both baseline models to cool, we find 
very similar results (over a large range of masses), in the sense that both 
give rise to extremely high cooled gas fractions and both predict $L-T$ 
relations which are too luminous (at a fixed temperature) with respect to the 
observations (we demonstrate this explicitly in \S 3 and \S 4 for the 
isothermal model).  {\it Therefore, we find that including the effects of 
radiative cooling but not those of entropy injection leads to failure in 
accounting for observed properties of clusters.  This conclusion holds 
irrespective of which baseline model we use.}  Throughout the paper, we 
present results for the isothermal plus cooling model only.
 
Perhaps a more physically plausible model is one which includes both the 
effects of radiative cooling and entropy injection (e.g., Voit et al. 2002, 
2003; Oh \& Benson 2003).  In order to examine this scenario, we will cool the 
entropy injection model of Babul et al. (2002).  Like the isothermal model, the 
entropy injection model also consists of intracluster gas in hydrostatic 
equilibrium within a realistic dark halo.  The primary difference between 
this model and the isothermal model (aside from the fact that one model assumes 
isothermality and the other does not) is that in the absence of cooling, the 
entropy injection model contains an isentropic core, which is presumed to have 
arisen through early heating events such as AGN outflows (e.g., Valageas \& 
Silk 1999; Babul et al. 2002; Scannapieco \& Oh 2004).  The value of 
the entropy of this core is a free parameter and has been determined 
previously by fitting to observed scaling relations.  Analysis of ``cooling 
flow corrected'' scaling relations, such as the $L-T$ relation (Babul et al. 
2002), $M_{\rm gas}-T$ relation (McCarthy et al. 2002a), and various SZ 
effect scaling relations (McCarthy et al. 2003b), indicates that an entropy 
core of $\gtrsim 300$ keV cm$^2$ gives the best fit.  It is interesting to see 
whether or not such a high level of injection is required once the effects of 
radiative cooling are also included and the results compared to actual 
uncorrected X-ray data.  Before moving on, it is also worth noting that the 
entropy profile at large radii (where entropy injection is unimportant) in 
this model is not identical to that of our baseline isothermal model.  Instead, 
the profile at large radii is required to match the results of high resolution 
non-radiative simulations (Lewis et al. 2000).  We verify that the slope and 
normalization of the entropy profile at large radii is also a close match to 
the self-similar entropy profile reported by Voit et al. (2003).

\subsection{A treatment of radiative cooling}

Voit et al. (2002) clearly demonstrated that how one chooses to model the effects 
of radiative cooling can have a significant impact on the predicted 
properties of clusters (compare the results of their `truncated' cooling model 
with their more realistic `radiative losses' cooling model, for example).  Thus, 
we wish to treat the effects of radiative cooling as realistically as possible 
but without resorting to computationally expensive hydrodynamic simulations.  
The treatment developed below is similar to the physically-motivated analytic 
method of Oh \& Benson (2003) and the reader is referred to that study for a 
more in-depth discussion of boundary conditions and how the relevant 
differential equations are solved.  We give a description of this method below.

We start with the initial gas and dark matter radial profiles for the model 
clusters (i.e., the profiles predicted by the isothermal and entropy injection 
models described above) and subject these to radiative cooling.  Radiative 
cooling reduces the specific entropy ($s$) of a parcel of gas according to

\begin{equation}
\frac{ds}{dt} = -\frac{\mu m_H n_i n_e \Lambda(T)}{\rho k_b T}
\end{equation}

\noindent where $\Lambda(T)$ is the cooling function (which is modeled as a 
Raymond-Smith plasma with 0.3 solar metallicity), $\mu$ is the mean molecular 
weight ($0.6$ in this case), and the other symbols have their usual meanings.

Equation (1) can be re-written in terms of the gas pressure ($P$) and the 
more commonly used form of `entropy' $K$, where $s = \ln{K^{3/2}} +$ constant 
(assuming an ideal gas and an equation of state $P = K \rho^{5/3}$),

\begin{equation}
\frac{dK}{dt} = -\frac{2}{3} \biggl(\frac{n_e n_i}{n^2} \biggr) \frac{1}{(\mu 
m_H)^2} \biggl(\frac{P}{K} \biggr)^{1/5} \Lambda(K,P)
\end{equation}

The new gas entropy profile after cooling for a small time interval $dt$ is 
calculated by integrating equation (2).  The gas pressure is assumed to 
remain constant (i.e., isobaric cooling) over this short interval.  After each 
time step, the properties of the gas (density and temperature and, therefore, 
pressure) are updated by placing the model clusters back in hydrostatic 
equilibrium by simultaneously solving the coupled differential equations

\begin{eqnarray}
\frac{dr}{dM_{\rm gas}} & = & \frac{1}{4 \pi r^2} 
\biggl(\frac{K}{P}\biggr)^{3/5} \nonumber\\
\frac{dP}{dM_{\rm gas}} & = & -\frac{G M_{\rm DM}}{4 \pi r^4}
\end{eqnarray}

As the gas cools and the pressure at the cluster center preferentially 
decreases, an inward flow develops in order to re-establish hydrostatic 
equilibrium.  Because the gas flows inward, we must implement different 
boundary conditions than employed in setting up the initial cluster profiles 
(i.e., $M_{\rm gas, tot}/M_{\rm tot} = \Omega_b/\Omega_m$ at $r_{\rm halo}$). 
We solve for hydrostatic equilibrium after each cooling time step by applying 
the following boundary conditions (see Oh \& Benson 2003):

\begin{eqnarray}
r(0) & = & 0 \nonumber\\
r(M_{\rm gas, tot}) & = & r_{\rm end} \nonumber\\
P(M_{\rm gas, tot}) & = & \biggl[P_{\rm halo}^{2/5} + \frac{2}{5 
K_{\rm halo}^{3/5}} \int^{r_{\rm halo}}_{r_{\rm end}} \frac{G M_{\rm 
DM}(r)}{r^2} dr \biggr]^{5/2}
\end{eqnarray}

\noindent where $P_{\rm halo}$ and $K_{\rm halo}$ are the initial gas 
pressure and entropy at $r_{\rm halo}$, the maximum radius of cluster (see 
Babul et al. 2002 for a quantitative definition of $r_{\rm halo}$).  The last 
boundary condition implies that the outermost gas mass shell is compressed 
adiabatically as it flows inward, which is appropriate since the cooling time 
of this shell greatly exceeds the age of the cluster.  Since this is a two 
point boundary value problem (with $r_{\rm end}$ being the eigenvalue of the 
problem), we use a relaxation technique to solve the equations.

If cooled long enough, the temperature of the gas at the center of the model 
clusters will approach zero and cease to emit X-rays.  This is referred to as 
`dropping out'.  For the purposes of the present model, we assume a parcel of 
gas drops out (and is then removed from the calculation) if its temperature 
falls below $\approx 10^{5}$ K or if its entropy decreases to zero during a 
time step.  As the gas at the cluster center approaches this threshold, the 
time steps are chosen such that only a few mass shells drop out at a time.  
Typically, this corresponds to a temporal resolution of 20 Myr (depending on 
system mass), which is small compared to the age of the cluster and is more 
than sufficient to achieve convergent results.  As the gas starts to drop out, 
a cooling flow is established.  This flow is treated as an adiabatic process 
after the small time step $dt$ by shifting the remaining (hot) gas, $K(M_{\rm 
gas})$, inward to replace the mass shells that dropped out.  The properties of 
the gas are then updated via hydrostatic equilibrium (as described above) and 
the cluster continues to cool.  Similar to Oh \& Benson (2003), we do not 
consider the effects of the cold gas on the gravitational potential of the 
cluster (e.g., adiabatic contraction of the dark matter halo).  However, we 
do not expect this to significantly modify our results as our entropy 
injection + cooling model generally predicts small cooled gas fractions (which 
are consistent with observations).

An obvious but important question is how long should cooling be allowed 
to operate?  For the sake of simplicity, it has become standard to allow the 
model clusters to cool for a Hubble time, $t_h$ (e.g., Bryan 2000; Voit \& Bryan 
2001; Voit et al. 2002; Xue \& Wu 2002; Oh \& Benson 2003).  Clearly, this 
represents the maximum amount of cooling a cluster can undergo.  We show later 
(in \S 4) that if one focuses solely on explaining relaxed CF clusters, that in 
fact a wide distribution of times (in addition to an entropy injection level of 
$\lesssim 300$ keV cm$^2$) is required in order to account for the scatter in 
the $L-T$ and $L-M$ relations.  The possibility of a connection between the 
intrinsic scatter in the $L-T$ relation and variations in the time available 
for cooling was previously suggested by Scharf \& Mushotzky (1997).  Throughout 
the paper, we plot results that span cooling from $t = 0$ to $t = t_h$.  

\section{The effects of radiative cooling}

We examine here the general effects of radiative cooling on the global and 
structural properties of isothermal and entropy injection model clusters.  This 
will aid the discussions in \S 4 and \S 5 of comparisons with the observations.

\subsection{Cooled gas fractions}

As discussed earlier, if a cluster is allowed to cool for a long enough time, 
eventually gas will drop out of the ionized X-ray emitting phase, become 
neutral, and possibly form stars.  Since there are fairly good observational 
constraints on the fraction of a cluster's baryons that are in the form of 
neutral gas (e.g., Donahue et al. 2000; Edge 2001; Edge et al. 2002) and stars 
(e.g., Cole et al. 2001; Lin, Mohr, \& Stanford 2003), a key prediction of 
the theoretical models is the fraction of gas that completely cools out.

Plotted in Figure 1 is the percentage of gas that completely cools out as a 
function of cluster mass, entropy injection level, and time (as the 
clusters cool).  The various line types demonstrate how the cooled fraction 
evolves as a function of time, while the different panels show how entropy 
injection affects the amount of gas that is able to completely cool.  The 
level of injection is characterized by $S$ (what X-ray observers often call 
the ``entropy''), which is related to the $K$ via

\begin{equation}
S \equiv \frac{k_b T}{n_e^{2/3}}= K \biggl(\frac{n}{n_e} \biggr)^{2/3} (\mu 
m_H)^{5/3}
\end{equation}

Concentrating for a moment on the upper left hand panel of Fig. 1, it can be seen 
that given nearly a Hubble time to cool, low mass clusters that have not been 
injected with entropy can cool out a substantial fraction ($\gtrsim 20$\%) of 
their baryons.  Yet, observations indicate that, at most, only 10\% of a 
cluster's baryons are in the form of stars (see Balogh et al. 2001).  A 
negligible amount is in the form of neutral gas (see Edge et al. 2002).  Simply 
reducing the amount of time that such systems can cool for (within reason) does 
not resolve this problem (see the dotted line, for example).  This predicted 
overabundance of cooled material in groups and clusters has been dubbed the 
``cooling crisis'' and has been taken as strong evidence in support of 
feedback/entropy injection (Balogh et al. 2001; Oh \& Benson 2003).  The 
requirement for large amounts of feedback/entropy injection also seems to be 
necessary in order to quench a similar problem found in semi-analytic and 
hydrodynamic studies of galaxy formation (Somerville \& Primack 1999; 
Yoshida et al. 2002; Benson et al. 2003).   

Encouragingly, the results for the isothermal plus cooling model are quite 
similar to those found from other analytic cluster models that include the 
effects of radiative cooling but not entropy injection (e.g., Voit et al. 2002; 
Oh \& Benson 2003).  This is despite there being slight differences in the 
adopted initial cluster conditions and
{\epsscale{1.2}
\plotone{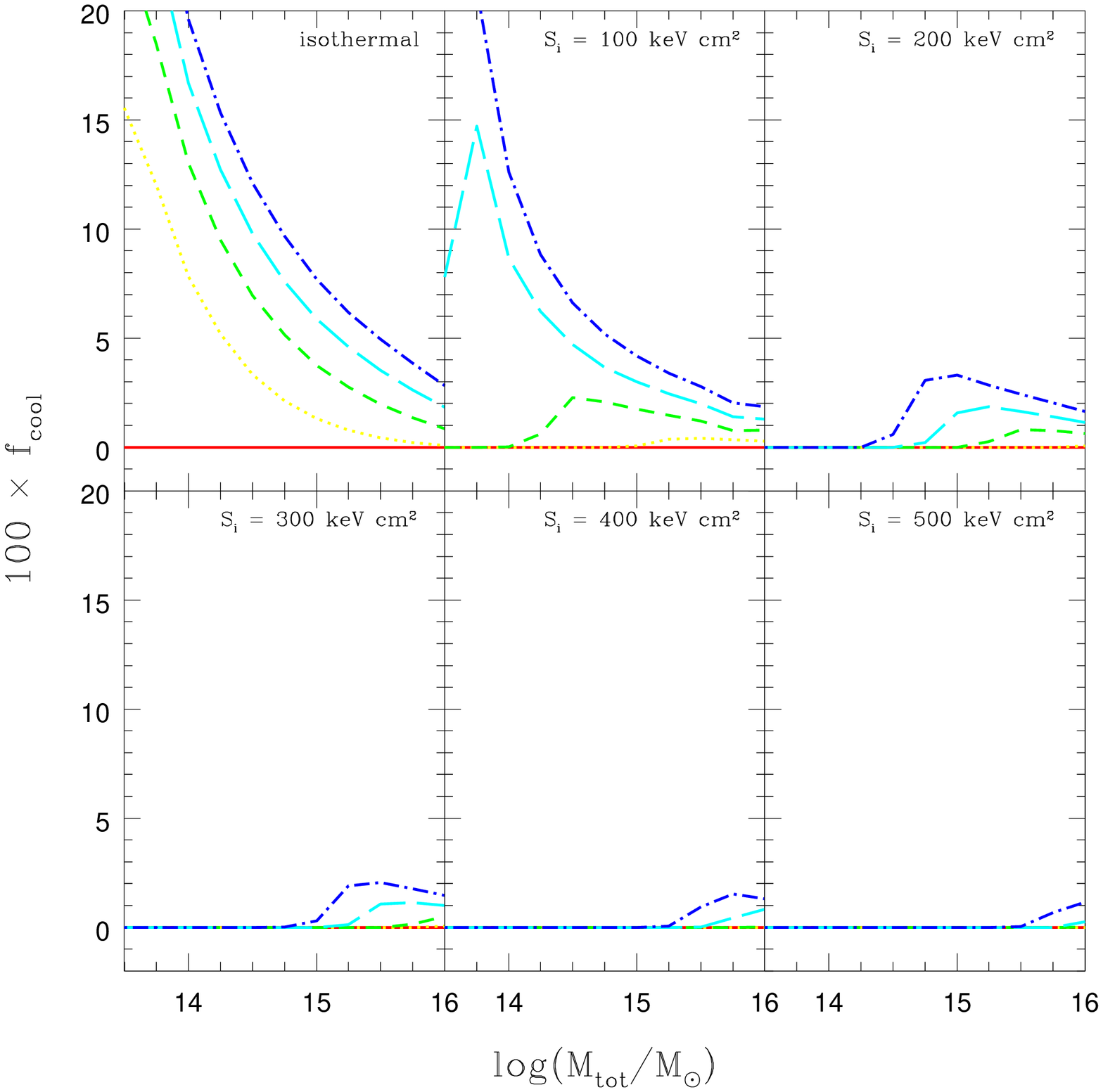}
{Fig. 1. \footnotesize
Percentage of the total gas mass that completely cools
out as a
function of total cluster mass, time, and entropy injection level $S_i$.  The
cooled gas fraction, $f_{\rm cool}$, is given by $M_{\rm cool}/(M_{\rm cool} +
M_{\rm hot})$, where $M_{\rm cool}$ is the total mass of gas that has cooled out
of the X-ray emitting phase and $M_{\rm hot}$ is the total mass of gas remaining
in the hot ionized phase.  The solid, dotted, short dashed, long dashed, and dot
dashed lines represent cooling for 0, 3, 6, 9, and 12 Gyrs, respectively.  The
different panels indicate how various levels of entropy injection affect the
amount of gas that is able to completely cool out.}}
\vskip0.1in
\noindent
how one chooses to model the effects of 
cooling.  This model also predicts cooled gas fractions which are comparable, 
although slightly lower, than those found from numerical simulations that include 
radiative cooling only (e.g., Muanwong et al. 2002; Dav\'{e}, Katz, \& Weinberg 
2000).  

Examination of the remaining panels in Fig. 1 clearly demonstrates that 
entropy injection has a large effect on the amount of gas that is able to cool 
out of the X-ray emitting phase.  In particular, injecting the gas with 
$S_i \gtrsim 200$ keV cm$^2$ is sufficient to obtain cooled gas fractions 
consistent with the observations, while injecting more than $300$ keV cm$^2$ 
essentially shuts off cooling in all but the most massive clusters.  Hence, 
entropy injection offers a viable solution to the so-called cooling crisis (see 
Oh \& Benson 2003 for a detailed discussion).  

\subsection{Entropy profiles}

Through hydrostatic equilibrium, the properties of the hot X-ray emitting gas at 
any particular time are determined entirely by the entropy distribution of the 
gas and the structure of the cluster's dark matter halo (see eqns. 3 and 4).  
Since radiative cooling modifies the cluster's entropy profile (see eqn. 2) it must 
also effect the cluster's gas density and temperature.  These, of course, set the 
cluster's appearance and dictate how efficiently the cluster can continue 
cooling.  Understanding how cooling modifies the entropy distribution of a 
cluster, therefore, is of paramount importance in understanding how it
influences the evolution of a cluster's global and structural
{\epsscale{1.2}
\plotone{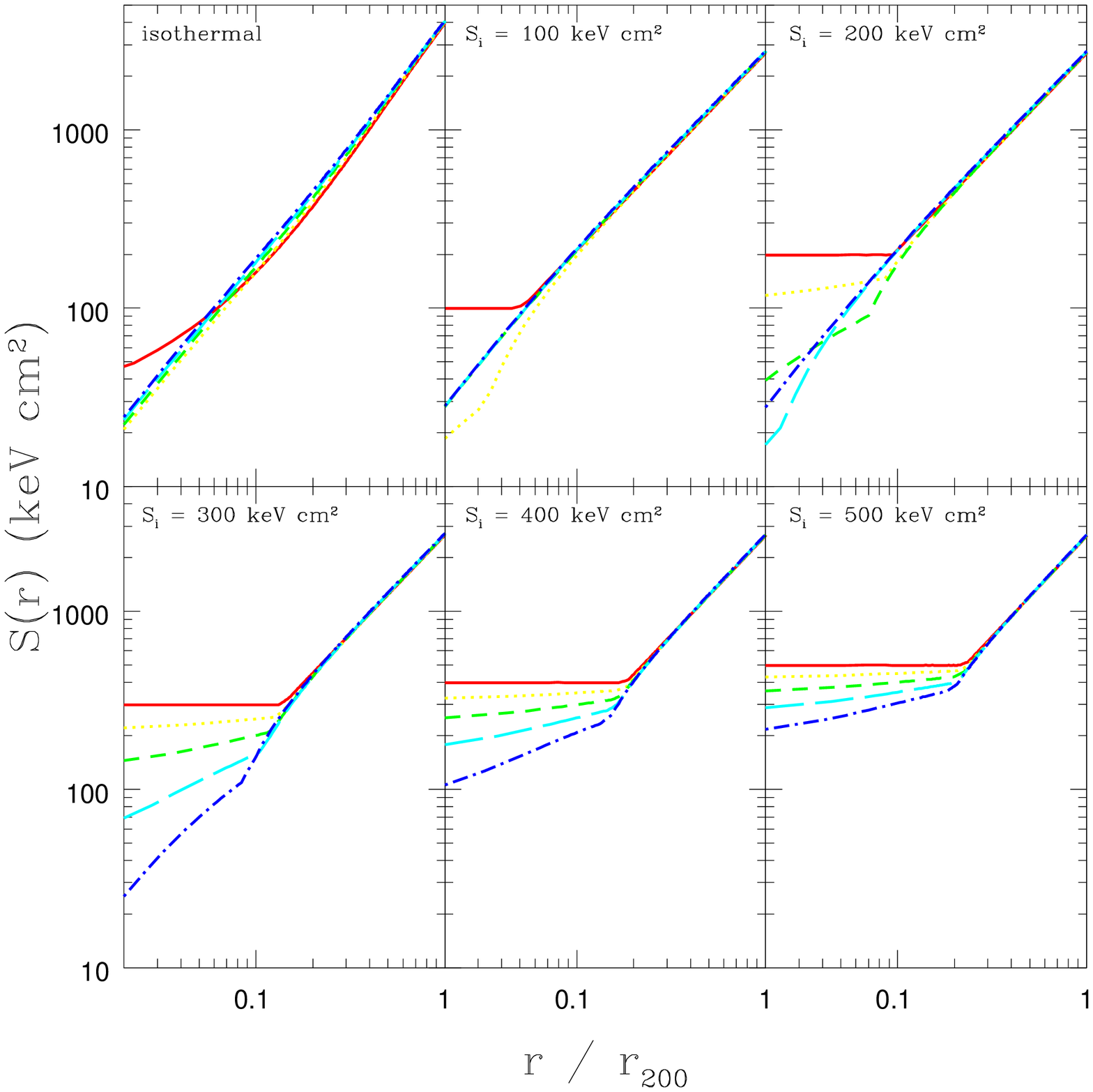}
{Fig. 2. \footnotesize
The entropy profile as a function of time and entropy injection
level for a cluster with $M_{\rm tot} = 10^{15}$ M$_{\odot}$.  The line types
have the same meaning as in Fig. 1.  The quantity $r_{200}$ is defined as the
radius within which the mean dark matter density is 200 times the critical
density of the universe (1.68 Mpc in this case).
}}
\vskip0.1in
\noindent
properties.

Figure 2 shows the evolution of the entropy profile of a cluster with $M_{\rm 
tot} = 10^{15} M_{\odot}$ as a function of time.  The various panels demonstrate 
how varying levels of entropy injection influence this evolution.

Let us focus first on the upper right hand panel of Fig. 2, since it shows the 
full range of entropy profiles as a function time.  The initial entropy profile 
(solid line) shows a central floor of $200$ keV cm$^2$ (by design) and a power 
law of $S \propto r^{\sim 1.1}$ at large radii, which matches the large radii 
results of semi-analytic smooth accretion models and high resolution 
non-radiative simulations (Lewis et al. 2000; Tozzi \& Norman 2001; Voit et al. 
2003).  After approximately 3 Gyr of radiative cooling, the central entropy of 
the cluster has dropped to nearly $100$ keV cm$^2$.  At this point, a clear 
entropy floor persists.  As the cluster continues to cool, the entropy core 
steepens until eventually the central entropy approaches zero.  When this 
occurs, gas begins to drop out of the X-ray emitting phase and an inward 
cooling flow develops.  However, even after 9 Gyr of cooling a remnant of the 
initial entropy is still present (note the kink in the entropy profile near 
0.03 $r_{200}$), although the `core' is now quite steep and very small in 
radial extent (since most of it has dropped out).  Eventually, the entropy core 
completely drops out and what remains is essentially a pure power law, $S 
\propto r^{\sim 1.1}$, that extends all the way from the cluster center to its 
periphery.  After this, the cluster continues to cool but approximately 
maintains this power law over all radii, reaching a quasi-steady state$^7$.

\footnotetext[7]{We use the phrase {\it quasi}-steady state instead of just
steady state since, although the entropy profile maintains the same shape and
normalization, it continues to decrease in radial extent with time as the
gas flows inward to re-establish hydrostatic equilibrium.}

The remaining panels in Fig. 2 are now easily interpreted.  Our baseline 
isothermal model, which has the lowest initial central entropy, starts cooling 
gas out the fastest and it is simple to see why this model predicts such large 
cooled gas fractions (Fig. 1).  Injecting the gas with $100$ keV cm$^2$ only 
slightly delays the development of a cooling flow.  Injection levels of $S_i = 
300$ keV cm$^2$ or higher, however, essentially prevent any gas from dropping 
out, although the central entropy of the gas is significantly lower after 
cooling for a Hubble time.

The above trends hold true for clusters with masses different than that 
considered in Fig 2. as well.  The only difference is the amount of time it takes 
for the entropy profile to evolve.  For example, because low mass clusters have 
lower central densities than high mass clusters (and, therefore, are less 
luminous), they have a much more difficult time in cooling out their entropy 
cores.  On the other hand, clusters more massive than the one considered in Fig. 
2 cool out their cores more quickly and reach the quasi-steady state 
faster.

The fact that radiative cooling approximately maintains the initial 
power-law entropy profile with time is interesting and deserves some 
investigation.  First, it is worth noting that both the isothermal and entropy 
injection models approximately maintain their power-laws (once the 
elevated entropy at the center cools out).  This is despite the fact that the 
power-law indices are not identical for these two cases.  In particular, at 
large radii, the entropy injection model initially has $S \propto r^{1.1}$ 
while the isothermal model initially has $S \propto \rho^{-2/3} \propto 
r^{4/3}$ ($\rho \propto r^{-2}$ at large radii for this model).  Therefore, the 
fact that the power-law remains essentially invariant with cooling 
appears to be independent of the power-law index.  We have verified that this is 
roughly true for a range of different initial power-law indices.  
Interestingly, a very similar trend has recently been reported by Kaiser \& 
Binney (2003).  These authors demonstrated that radiative cooling does not 
significantly modify the initial power-law relationship between entropy and gas 
mass of their model clusters.  Unfortunately, a straightforward analytic 
explanation for these (numerically-derived) trends is not easily obtained, at 
least at small radii.  At large radii, however, we should expect the power-law 
to be maintained since the cooling time of the gas is long relative to the 
Hubble time.  This implies that $S(M_{\rm gas})$ should remain roughly constant 
and, furthermore, the physical size of a mass shell should be fixed (since the 
shell will not have been compressed much).  Understanding the evolution of the 
entropy profile at small radii is more difficult because both coordinates, 
$S(M_{\rm gas})$ and $r(M_{\rm gas})$, are being significantly modified by 
cooling.  A more thorough investigation of entropy evolution of clusters will 
be presented in a forthcoming paper.

Given enough time, our models predict that all clusters should reach a 
quasi-steady state that is characterized by a near perfect power law 
entropy profile over all radii.  However, since the age of the universe (which is 
an upper limit on the amount time available for cooling) is comparable to the 
predicted central cooling times of our model clusters, a general prediction of 
our model is that there should be a full range of central entropy 
distributions, depending on the initial injection level and how long each 
cluster is able to cool.  Indeed, in \S 5 we demonstrate that 
{\epsscale{1.2}
\plotone{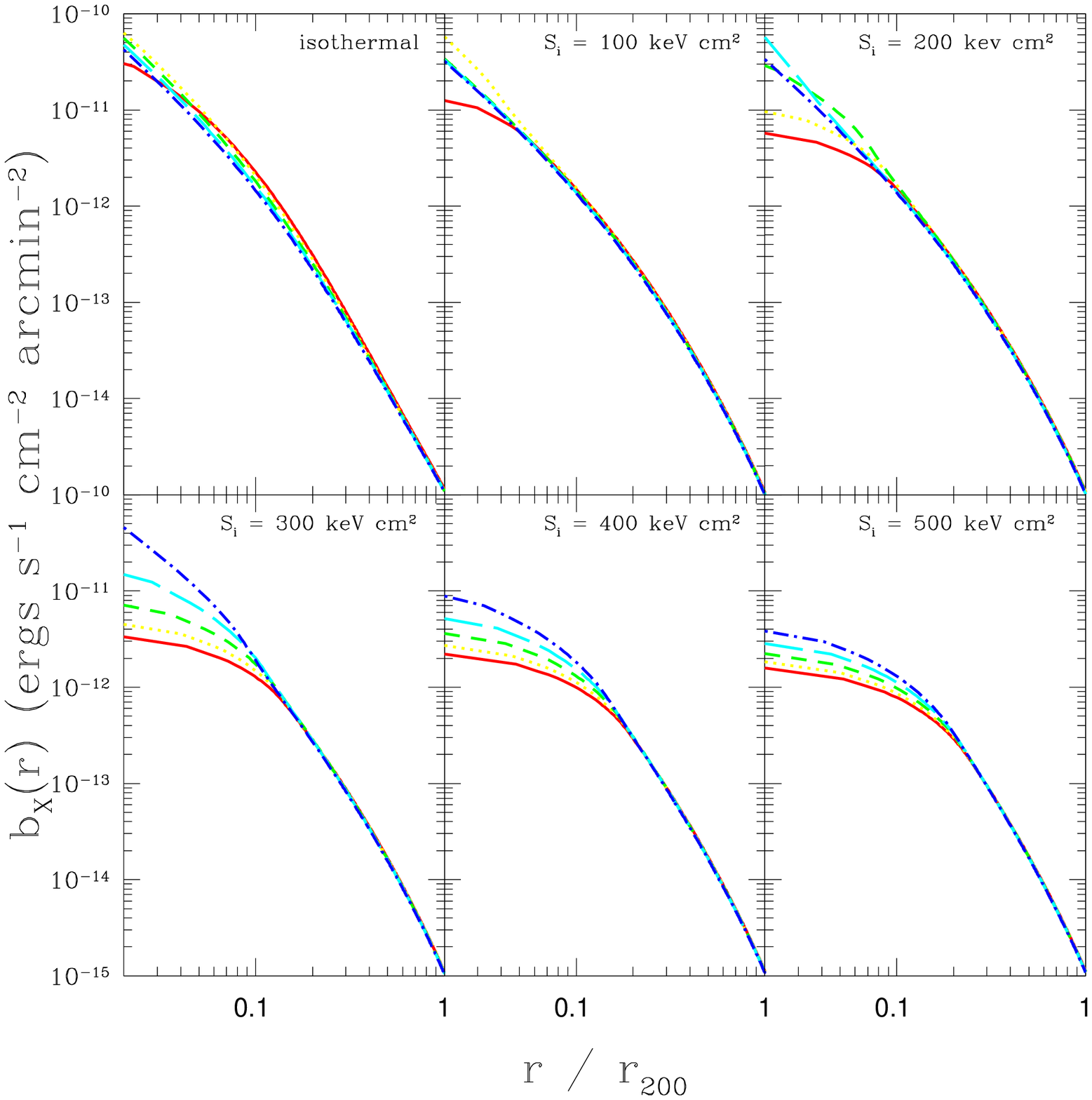}
{Fig. 3. \footnotesize
The bolometric X-ray surface brightness profile as a function of
time and entropy injection level for a cluster with $M_{\rm tot} = 10^{15}$
M$_{\odot}$. The line types have the same meaning as in Fig. 1.
}}
\vskip0.1in
\noindent
new published 
{\it Chandra} and {\it XMM-Newton} results show a large range of central 
entropies that compares quite favorably to those plotted in Fig. 2.

\subsection{Surface brightness and emission-weighted temperature profiles}

Plotted in Figure 3 is the evolution of the bolometric X-ray surface 
brightness ($b_X$) profile of a cluster with $M_{\rm tot} = 10^{15} 
M_{\odot}$ as a function of time.  The panels and line types have the same 
meaning as in Figs. 1 \& 2.

Again, we focus first on the model with an injection level of $S_i = 200$ keV 
cm$^2$ (upper right hand panel).  It can clearly be seen that as the cluster 
cools the surface brightness near the center of cluster increases and becomes 
more peaked.  This, of course, is due to the increasing central density 
which, in turn, is the result of the decreasing central entropy and the 
re-adjustment of the cluster gas to achieve hydrostatic equilibrium.  Thus, 
there is a strong connection between the evolution of the surface brightness 
profile and the evolution of the entropy profile of a cluster.  Indeed, 
comparison of Fig. 3 with Fig. 2 demonstrates that the central surface 
brightness is a strong function of the amount of low entropy gas near the 
cluster center.  For example, the central surface brightness reaches its 
maximum value after roughly 9 Gyr of cooling (long dashed line), which 
coincides exactly with the time when the central entropy reaches its lowest 
value (i.e., just before the core drops out).  Once the entropy core 
disappears, the surface brightness profile achieves a quasi-steady state 
(dot dashed line).  This is expected given the results of \S 3.2.  The same 
is true for other levels of entropy injection, the only difference being the 
time it takes to achieve this quasi-steady state (which are greater than a 
Hubble time for sufficiently high levels of entropy injection).  

The trend of increasing central concentration of $b_X$ as 
{\epsscale{1.2}
\plotone{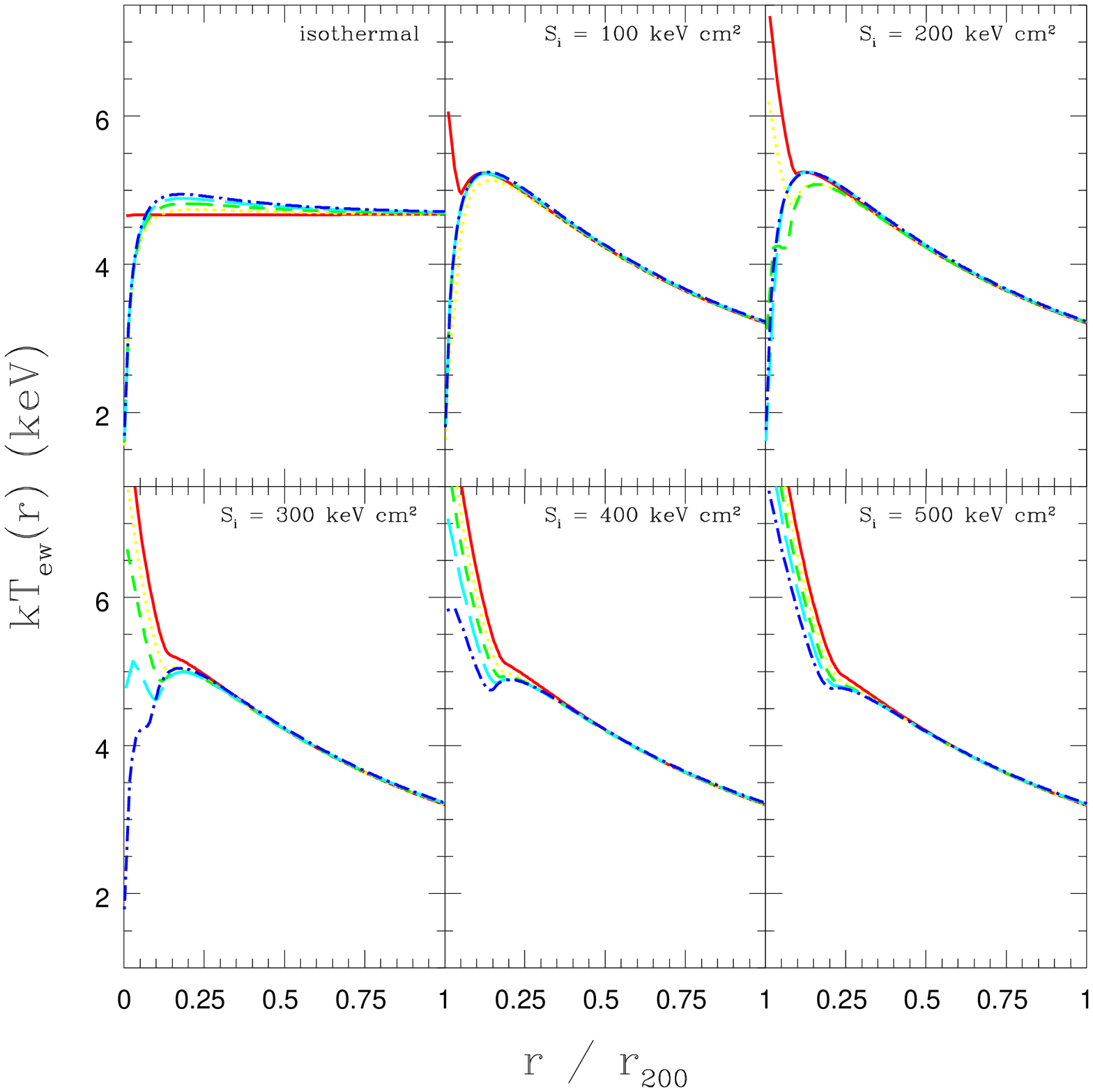}
{Fig. 4. \footnotesize
The emission-weighted temperature profile as a function of time and
entropy injection level for a cluster with $M_{\rm tot} = 10^{15}$ M$_{\odot}$.
The line types have the same meaning as in Fig. 1.
}}
\vskip0.1in
\noindent
the cluster cools is 
an interesting one.  For clusters that had a mild level of entropy 
injection (top three panels of Fig. 3), radiative cooling has a large effect 
on the central gradient of the cluster's surface brightness.  On the other
hand, clusters that had a large injection of entropy are less affected 
by cooling, as expected.  This difference in central concentration means we can
qualitatively identify the two types of clusters (i.e. those with mild and
strong heating) with CF and NCF clusters, respectively (see \S 5.2).  
In fact, we argue later, on the basis of the observed $L-T$ and $L-M$ 
relations, that the 
origin of these two morphological classes of clusters can be explained in terms 
of the entropy injection level.  This hypothesis is reinforced by the actual 
observed entropy profiles of these two classes of systems, which is presented 
in \S 5.

In Figure 4 we plot the evolution of the bolometric emission-weighted 
temperature ($kT_{\rm ew}$) profile of a cluster with $M_{\rm tot} = 10^{15} 
M_{\odot}$ as a function time.  Again, the panels and line types have the same 
meaning as in the previous figures.

In the top three panels of Fig. 4, we see the rapid development of large 
positive temperature gradients at the cluster center.  Like the surface 
brightness and entropy profiles discussed above, the emission-weighted 
temperature profile also reaches a quasi-steady state once the initial entropy 
core has dropped out.  The predicted steady state temperature profile, which is 
characterized by a steep positive gradient at the cluster center and a gentle 
negative gradient at large radii, is qualitatively similar to that recently 
observed in CF clusters by Allen, Schmidt, \& Fabian (2001) and De Grandi \& 
Molendi (2002).  On the other hand, clusters that had a high level 
of entropy injection ($S_i > 300$ keV cm$^2$) retain their sharp central 
negative temperature gradients even after cooling for more than 12 Gyr.
  
Lastly, it is also worth noting that the development of central positive 
temperature gradients in our models coincides almost exactly with the 
development of peaked surface brightness profiles.  In other words, depending on 
the time elapsed since the entropy injection, one either has a cluster with a 
flat central surface brightness and a sharp central negative temperature 
gradient or a cluster with a peaked surface brightness profile and 
a central positive temperature gradient.  These two types of clusters, which 
emerge naturally from our analytic model, match the main qualitative 
features of observed relaxed NCF and CF clusters.

\subsection{Integrated luminosities and mean cluster temperatures}

Figures 3 and 4 illustrate how radiative cooling is expected to modify the 
observable radial properties of the intracluster medium.  Here, we describe the 
effects of cooling on the integrated X-ray luminosity and the mean 
emission-weighted temperature of the cluster.

As is evident from Figs. 2-4, cooling primarily affects only the central regions 
of the model clusters.  Therefore, since cooling increases the central surface 
brightness and decreases the central temperature, we should expect that, in 
general, the integrated luminosity of the clusters will increase with the 
addition of radiative cooling while the mean temperature should decrease.  This 
is exactly what the model predicts, as shown by the evolutionary $L-T$ tracks 
plotted in Fig. 5.  Here, the tracks of six clusters of varying mass are 
shown (dashed and dotted lines, see figure caption).  The clusters evolve from 
the initial $L-T$ relation predicted by the $S_i = 200$ keV cm$^2$ model 
without cooling (thick solid line) to higher luminosities.  The open pentagons 
represent specific times of 0, 3, 6, 9, and 12 Gyr during the evolution.

To understand what this plot is telling us, let us focus for a moment on the 
tracks for the least massive (track A) and second most massive (track 
E) clusters.  Track A shows that the least massive cluster starts off with a 
luminosity of $L_{\rm X,bol} \approx 6 \times 10^{43}$ ergs s$^{-1}$ and 
evolves to higher luminosities.  After approximately 13 Gyr of cooling (i.e., 
the end of the track), the cluster's luminosity has increased by roughly a 
factor of 3.3.  Also, its temperature has decreased but not by nearly as much 
(only a factor of 1.19).  Thus, the evolution in the $L-T$ plane is primarily 
driven by the influence of cooling on the luminosity rather than the 
temperature.  It is important to note that 
even after 13 Gyr of cooling this low mass cluster has been unable to rid 
itself of its initial entropy core (see Fig. 1).  Therefore, it is 
straightforward to understand why the $L-T$ relation evolves as it does: with 
time the central entropy decreases, giving rise to an increased central density 
which, in turn, dramatically increases the X-ray luminosity of the cluster.  
More massive clusters experience something slightly different.  Focusing on 
track E, we see that just shortly after 6 Gyr of cooling the cluster begins to 
evolve back towards lower luminosities and higher temperatures (as indicated by 
the dotted line).  Eventually, after approximately 9 Gyr, it reaches a more or 
less stable position on the $L-T$ diagram.  These trends can be understood by 
re-examining the results presented in \S 3.3.  First, the fact that the cluster 
reaches a stable position on the $L-T$ diagram is expected since it has been 
shown that the surface brightness and temperature profiles reach quasi-steady 
states.  
{\epsscale{1.2}
\plotone{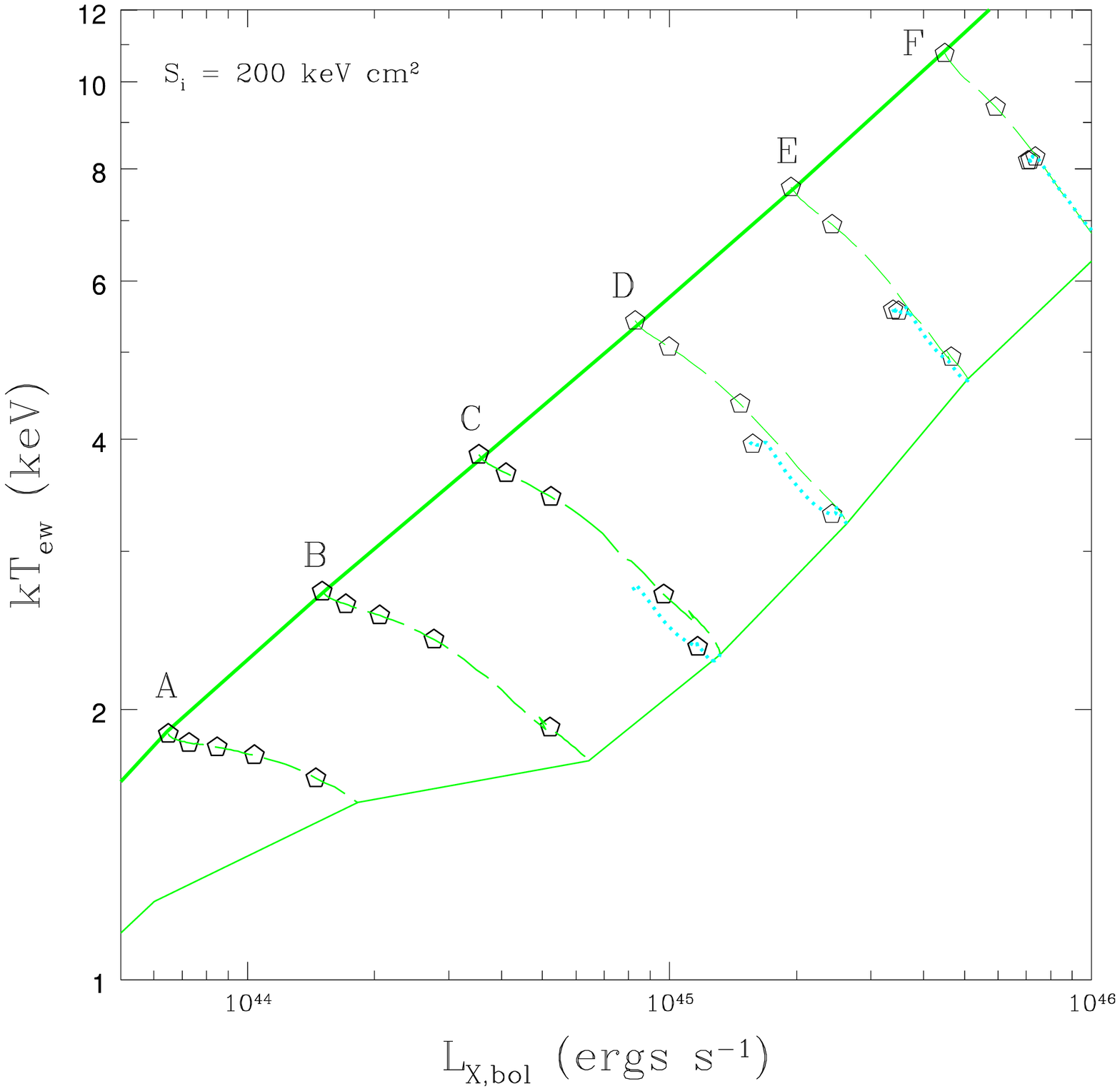}
{Fig. 5. \footnotesize
Evolutionary $L-T$ tracks for clusters that have been injected with
$S_i = 200$ keV cm$^2$.  Six different tracks are shown corresponding to
clusters with masses of $log_{10}(M_{\rm tot}) =$ 14.25 (A), 14.50 (B), 14.75
(C), 15.00 (D), 15.25 (E), and 15.50 (F) (in $M_{\odot}$).  The thick solid line
is the initial $L-T$ relation prior to including the effects of radiative
cooling.  The dashed and dotted lines are the evolutionary tracks as the cluster
cools.  The dashed lines show evolution towards high $L$ and low $T$, which
occurs prior to the removal of the isentropic core, while the dotted lines show
the evolution towards low $L$ and high $T$, which occurs after the core has
dropped out (if it is able to drop out in less than 13 Gyr).  Together, the thin
and thick solid lines enclose the predicted range of $L-T$ values during cooling
for 13 Gyr.  For comparison with the previous figures, the open pentagons show
the predicted $L-T$ relation at the discrete times of 0, 3, 6, 9, and 12 Gyr.
}}
\vskip0.1in
\noindent
Given the results of \S 3.3, it is also clear why this stable point 
happens to lie at a higher luminosity/lower temperature than the initial 
$L-T$ relation (i.e., the surface brightness is more peaked while the temperature 
profile shows a central positive gradient).  Furthermore, we can also understand 
the origin of the turn-around of this massive cluster on the $L-T$ diagram as 
being due to the entropy core in its final days.  Just prior to the dropping 
out of the entropy core, there exists a large amount of low entropy gas near 
the cluster center (since the core cools somewhat intact, see top right hand 
panel of Fig. 2).  This gives rise to an extremely peaked surface brightness 
profile.  However, once the core completely drops out, the surface brightness 
peak is diminished (so, too, is the integrated luminosity) and, hence, the 
result is the $L-T$ turn-around.  

As might be expected, the above trends also hold true for clusters that have 
experienced different levels of entropy injection.  Even clusters that 
were initially isothermal (and have no core per se) evolve towards higher 
luminosities and lower temperatures.  This seems to be a very general 
prediction of our cooling model for massive clusters.  Encouragingly, Voit et 
al.'s (2002) `radiative losses' model, which is similar to our model in many 
aspects, also predicts an overall evolution of the $L-T$ relation towards 
higher luminosities and lower temperatures and by roughly the same magnitude 
(see their Fig. 27).

Figure 5 shows that the position of any given cluster on the L-T diagram is 
dictated by the initial level of entropy injection and how long radiative 
cooling has had to operate. (See also Figure 9 for predicted L-T relation for a 
range of entropy injection levels.)  This implies a potential degeneracy in the 
sense that one can account for the observed location of any individual cluster 
on the $L-T$ diagram as being due to either a high level of initial entropy 
injection and a long subsequent period of cooling, or a low level of initial 
entropy injection followed by a relatively short period of cooling.  However, 
this degeneracy is bounded.  There are regions of the $L-T$ plot (see Figure 9) 
that cannot be accessed by clusters with $K_0 > 300$ keV cm$^2$, even if they 
cooled for the entire lifetime of the Universe.  Similarly, there are regions of 
the $L-T$ plot that cannot be accessed by clusters with low initial entropy 
injection.  Jointly, these two bounds can be used to place rough constraints on 
the initial entropy injection of an individual cluster.  On the other hand, as 
we discuss below, the distributions of present-day global and structural 
properties of the cluster population, collectively, provide a powerful probe of 
the initial {\it distribution} of entropy injection levels.  Additional insights
are also likely to be obtained from an analysis of a high-$z$ cluster sample 
because radiative cooling will not have had much time to operate in these 
systems.

Before proceeding with a comparison of our models with the observational data, 
we stress that there is an important caveat to the results presented in this 
subsection.  
As noted earlier, if a cluster is able to cool out its initial entropy core, 
what remains is pure power law entropy profile with the cluster achieving a 
quasi-steady state.  This gives rise to a more or less fixed position on 
the $L-T$ diagram.  This is only true, however, if the radial extent of the 
entropy profile is not reduced by a significant amount after a Hubble time of 
cooling.  Under the circumstances considered above, i.e., high mass clusters 
and fairly large amounts of entropy injection, this condition is approximately 
met.  Because such clusters are only able to cool out a small fraction of their 
baryons, gas that was originally at the outskirts of the cluster only manages 
to flow inward by a relatively small amount.  However, in the limit of low 
mass clusters that experience small amounts of entropy injection (and, 
therefore, extremely high cooled gas fractions), the shrinking entropy profile 
can significantly affect the evolution of the $L-T$ relation.  The amount of 
gas that remains in the hot X-ray emitting phase after a Hubble time of cooling 
is reduced by a large fraction and so, too, is the X-ray luminosity of 
the cluster.  The result is a position on the $L-T$ diagram that is actually at 
a lower luminosity and higher temperature than the initial position (i.e., to 
the left of the initial curve).  In fact, the same would be true of high mass 
clusters if they could somehow be forced to cool out $\gtrsim 15-20$\% of 
their baryons.  We have tested this by allowing our high mass clusters to 
continue cooling for several Hubble times until they reach this cooled gas 
fraction.  Thus, the resulting $L-T$ relation is a strong function of the 
efficiency of cooling in clusters.  It is, therefore, essential that the model 
retain consistency with the observed cooled gas fraction of clusters. 

\section{Comparison with Observed Global Properties}

The luminosity-temperature relation has been known for roughly a decade now to 
deviate from simple self-similar predictions.  This has been the primary 
motivating factor for the investigation of cluster models that incorporate 
non-gravitational gas physics, such as entropy injection and/or radiative 
cooling.  While such models have generally been shown to provide better 
matches to the `typical' cluster, the large amount of intrinsic scatter in 
the observed $L-T$ (e.g., Novicki, Sornig, \& Henry 2002) and $L-M$ (e.g., 
Reiprich \& B\"{o}hringer 2002) relations has yet to be addressed by these 
models.  A primary goal of the present study is to see whether or not the 
scatter is consistent with theoretical models that include entropy injection 
and/or radiative cooling.  

\subsection{Observations}

Essential to exploring the properties of the scatter are large, homogeneously 
analysed samples of clusters.  Such samples guard against selection effects, 
differences in analysis procedures, differences in absolute flux 
calibration between various instruments, and other issues which 
could influence the physical interpretation of the scatter.  Ideally, these 
samples should not have had any corrections applied to them (e.g., ``cooling 
flow'' correction) other than the removal of obvious point sources (e.g., 
McCarthy, West, \& Welch 2002b).  Fortunately, such samples are now becoming 
available.  We focus on two of the larger samples that now exist: the {\it 
ASCA} Cluster Catalog (ACC) of Horner (2001) and the extended {\it ROSAT} 
HIFLUGCS sample of Reiprich \& B\"{o}hringer (2002). 

The ACC of Horner (2001) contains roughly 270 clusters, the large majority of 
which are nearby systems ($z \leq 0.2$) (and, henceforth, we restrict 
ourselves to nearby clusters with $z \leq 0.2$).  The wide bandpass and good 
spectral 
resolution of {\it ASCA} allows for the accurate determination of cluster 
mean temperatures and total bolometric X-ray luminosities.  Unfortunately, 
because of {\it ASCA}'s rather poor spatial resolution, it is not possible to 
accurately measure surface brightness profiles and, therefore, quantities such 
as cluster mass.  Thus, we use the ACC by itself to study the 
luminosity-temperature relation only.  In order to study the luminosity-mass 
relation of clusters, we turn to the extended HIFLUGCS sample.  Using surface 
brightness profiles from {\it ROSAT} and temperatures from {\it ASCA}, Reiprich 
\& B\"{o}hringer (2002) deduced, through the assumption of hydrostatic 
equilibrium (and isothermality), the masses of 106 nearby ($z \lesssim 0.2$) 
clusters within three different radii: $r_{500}$, $r_{200}$, and a fixed 
physical radius of $3 h^{-1}_{50}$ Mpc.  In the present study, we focus only 
on the mass within $r_{500}$, which is typically the smallest of the three 
and, therefore, requires the least amount of extrapolation of the observed 
surface brightnesses (see Reiprich \& B\"{o}hringer 2002).  Unfortunately, 
many of the temperatures used by these authors were corrected for the effects 
of ``cooling flows'' and, therefore, in its current state, the sample cannot be 
fairly compared to our radiative cooling models.  In order to address this 
issue, we have selected clusters in common between the ACC and the HIFLUGCS 
($\approx 80$ systems).  Using uncorrected temperatures from the ACC and 
surface brightness profiles from the HIFLUGCS, we recompute the cluster masses 
within 
{\epsscale{1.2}
\plotone{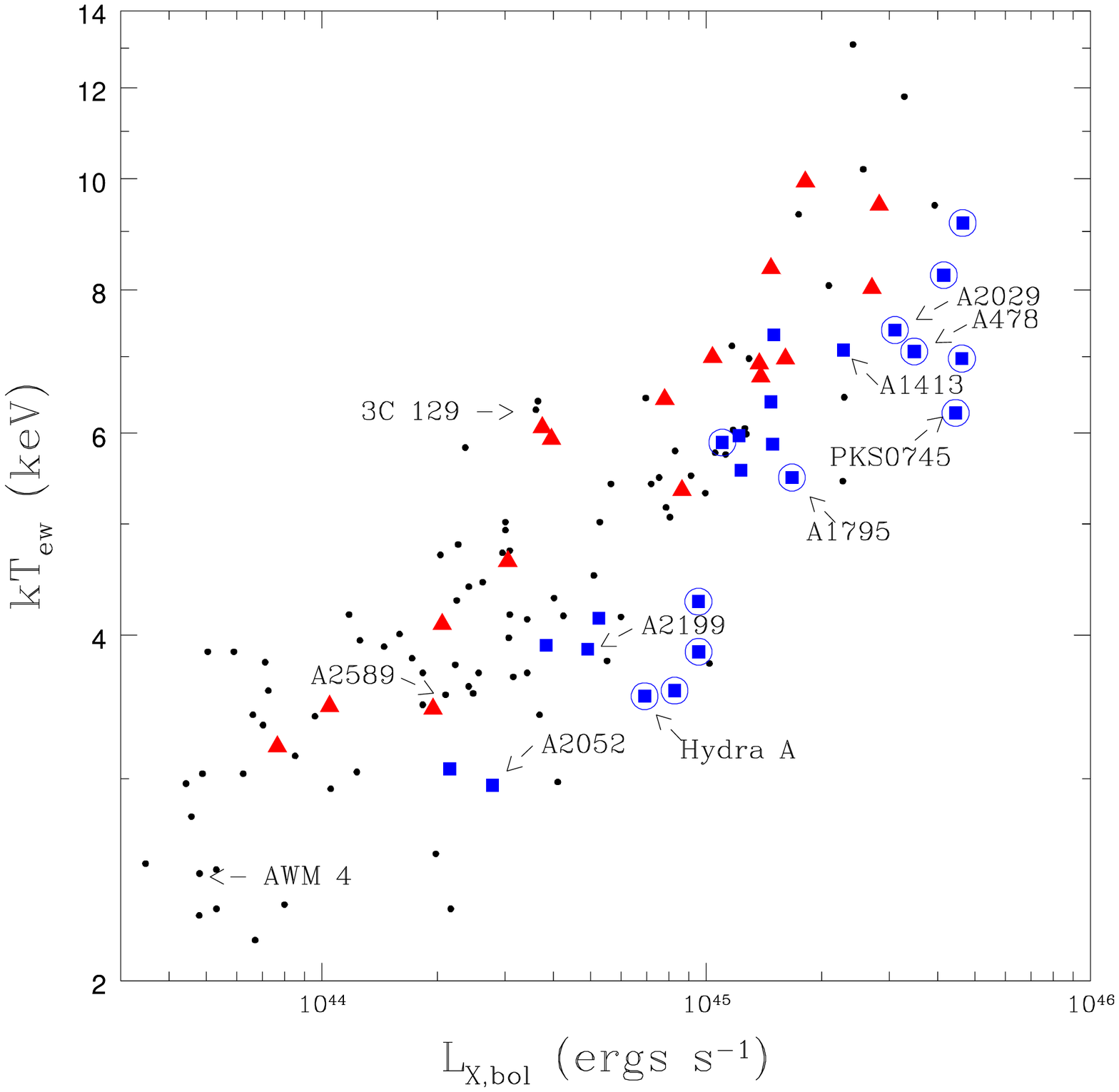}
{Fig. 6. \footnotesize
The observed $L-T$ relation of nearby, massive galaxy clusters.
Data are from the ASCA Cluster Catalog of Horner (2001).  Small filled circles
represent clusters that have no published cooling flow status.  Triangles
represent clusters that have an inferred cooling flow mass deposition rate,
$\dot{M}$, consistent with zero, squares represent clusters with $\dot{M}
\gtrsim 100$ $M_{\odot}$ yr$^{-1}$, and squares with a surrounding open circle
represent clusters with $\dot{M} \gtrsim 300$ $M_{\odot}$ yr$^{-1}$.  Cooling
flow deposition rates were estimated by Allen \& Fabian (1998) and Peres et al.
(1998) based on a deprojection analysis of {\it ROSAT} data.  The clusters that
have been labeled are those which now have published entropy profiles inferred
from {\it Chandra} or {\it XMM-Newton} observations.
}}
\vskip0.1in
\noindent
$r_{500}$, taking care to correct for differences in the assumed 
cosmologies.

Before comparing the observed scaling relations to the models, it is instructive 
to examine the properties of the data itself.  Plotted in Figure 6 is the 
observed luminosity-temperature relation of nearby, massive clusters.  Where 
possible, we indicate with symbols the cooling flow status of the clusters.  In 
particular, triangles represent clusters that have an inferred cooling flow 
mass deposition rate, $\dot{M}$, consistent with zero, squares represent 
clusters with $\dot{M} \gtrsim 100$ $M_{\odot}$ yr$^{-1}$, and squares 
with a surrounding open circle represent clusters with $\dot{M} \gtrsim 300$ 
$M_{\odot}$ yr$^{-1}$.  Cooling flow deposition rates were estimated by Allen \& 
Fabian (1998) and Peres et al. (1998) based on a deprojection analysis of {\it 
ROSAT} surface brightness data.  Those clusters that have published deprojected 
entropy profiles (or deprojected temperature and density profiles from which an 
entropy profile may be derived) inferred from new {\it Chandra} or {\it 
XMM-Newton} data have been explicitly labeled and will be examined in detail in 
\S 5.

First, there is the expected well-defined correlation between a cluster's 
luminosity and temperature.  However, there is a large amount of scatter in the 
plot.  As discussed by Allen \& Fabian (1998), correcting for the central 
``cooling flow'' dramatically reduces the scatter in the $L-T$ relation.  Here, 
we have purposely left the 
{\epsscale{1.2}
\plotone{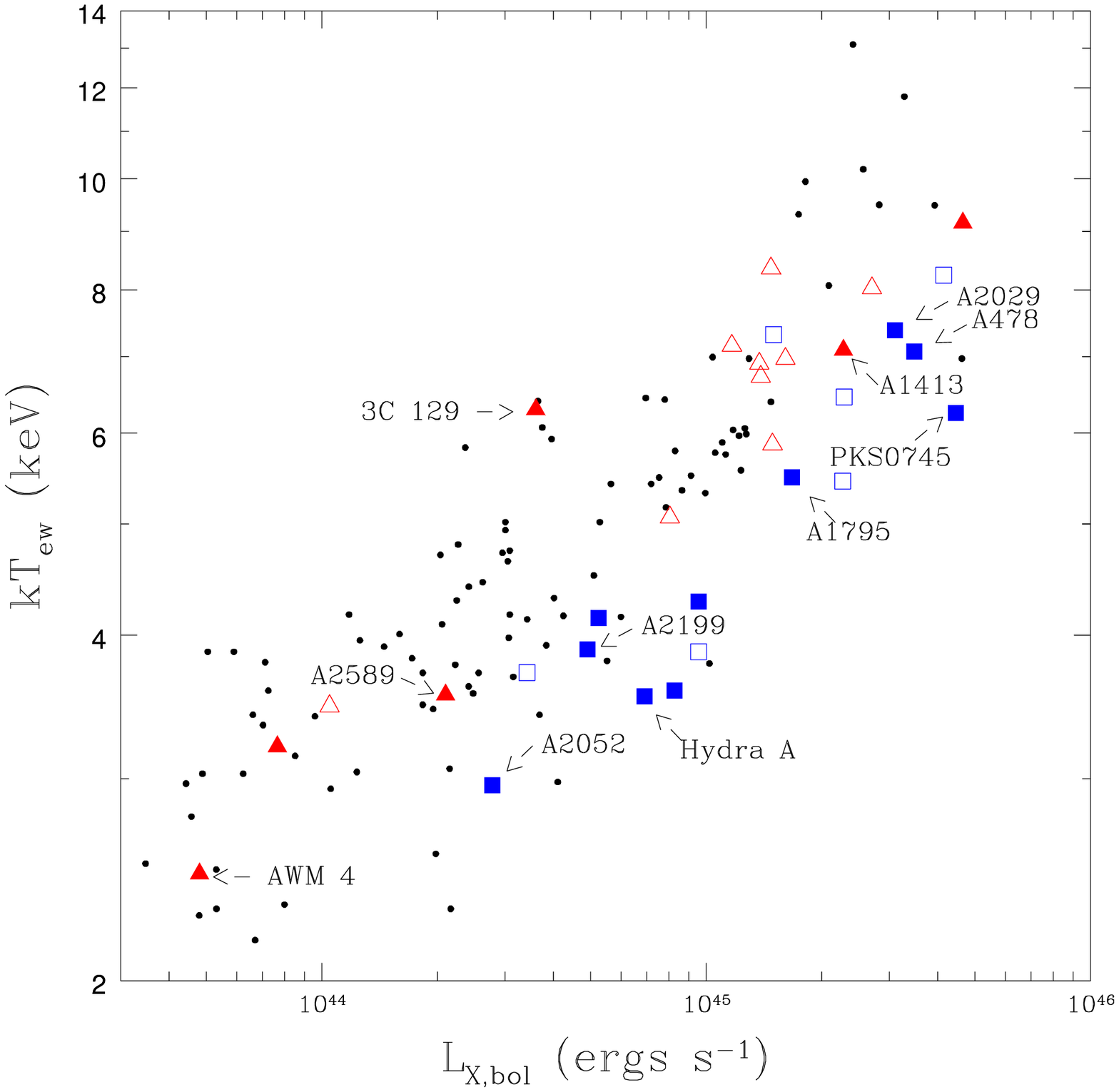}
{Fig. 7. \footnotesize
Same as Fig. 6 but using published {\it Chandra} and {\it
XMM-Newton} data to classify the clusters according to CF status and morphology.
Here, squares represent CF clusters, triangles represent NCF clusters, filled
symbols represent relaxed systems, and open symbols represent unrelaxed systems.
}}
\vskip0.1in
\noindent
data uncorrected for the effects of ``cooling 
flows'' in an attempt to ascertain whether models that include radiative 
cooling (either alone or in addition to some amount of entropy injection) can 
account for this scatter.

There is also a high degree of coherent structure present in the scatter of
the $L-T$ relation.  Namely, the clusters with large values of $\dot{M}$
preferentially lie on the high luminosity side of the scaling relation, 
while clusters with small values of $\dot{M}$ lie on the low luminosity side.  
This correlation between inferred cooling flow strength and dispersion in the 
scaling relation has been known for some time (Fabian et al. 1994; see also 
White, Jones \& Forman 1997; Allen \& Fabian 1998; Markevitch 1998).  However, 
Fig. 6 is probably the cleanest, most clearcut illustration of this trend to 
date, with the high and low $\dot{M}$ systems occupying well-defined locii.

Recently, the standard isobaric cooling flow model has been shown to provide 
a poor description of {\it XMM-Newton} spectra of CF clusters (e.g., Peterson et 
al. 2001, 2003; Kaastra et al. 2004).  Thus, the inferred mass deposition rates 
may not be perfect indicators of whether a cluster is a ``cooling flow'' cluster 
or not (i.e., whether or not it contains a peaked surface brightness and a 
central positive temperature gradient).  This is important if we wish to quantify 
comparisons between the models and CF or NCF clusters.  In order to test this 
idea, we searched the literature for new {\it Chandra} and {\it XMM-Newton} 
observations of clusters in the Horner ACC and classified each system as either 
CF or NCF, and either relaxed or unrelaxed.  Determination of CF versus NCF 
status is based primarily on the presence or absence of a well-defined central 
positive temperature gradient, while determination of the dynamical state is 
determined by the presence or absence of large-scale ($\sim$ a few 
hundred 
{\epsscale{1.2}
\plotone{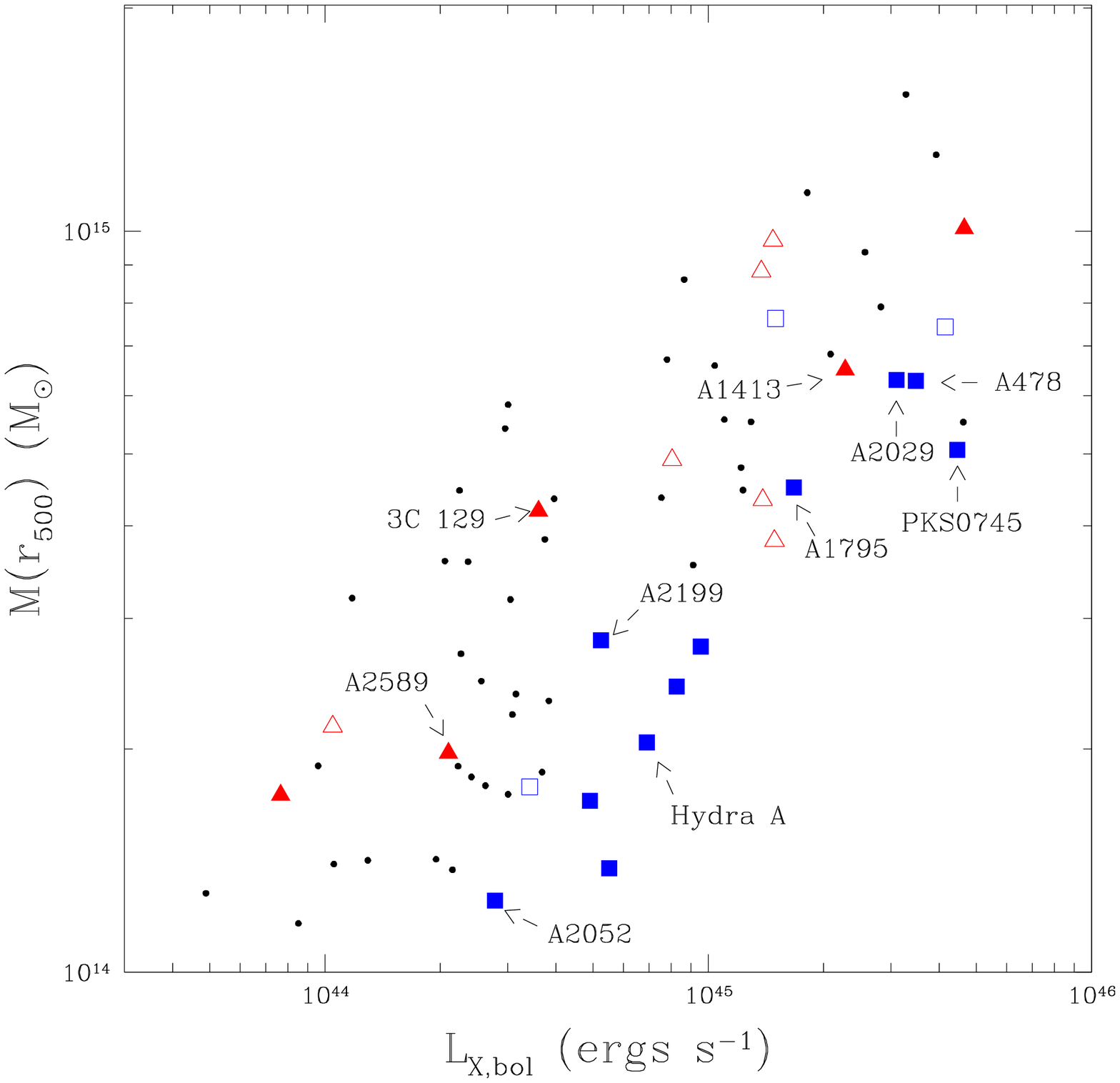}
{Fig. 8. \footnotesize
The observed $L-M$ relation of nearby, massive galaxy clusters.
Luminosities are from Horner (2001) and masses are computed using the surface
brightness profiles of Reiprich \& B\"{o}hringer (2002) and uncorrected
temperatures of Horner (2001).  The symbols have the same meaning as in Fig. 7.
}}
\vskip0.1in
\noindent
kpc) substructure in the X-ray images, which is presumably due to 
mergers.  {\it Therefore, our classification scheme makes use of observed 
features rather than quantities inferred by fitting an assumed model}.  The 
results of this classification are presented in Figure 7 for the $L-T$ 
relation and Figure 8 for the $L-M$ relation.  Table 1 lists the 
clusters plotted in Figs. 7 and 8 along with their 
classifications.

An examination of Fig. 7 demonstrates that our ``cooling flow'' classification 
scheme correlates well with the results plotted in Fig. 6, although the results 
in Fig. 7 look less impressive since, to date, only a relatively small number of 
clusters in the ACC have been observed by {\it Chandra} or {\it XMM-Newton}.
Furthermore, Fig. 7 indicates that there are apparently more unrelaxed NCF 
systems than relaxed systems.  It is important to note, however, that the 
majority of NCF clusters observed with {\it Chandra} to date have been selected 
because they were known to be mergers (for the purposes of studying bow shocks, 
etc.).  This is discussed further in \S 6 \& 7.  Interestingly, several of the 
clusters that lie roughly on the boundary between high and low $\dot{M}$ 
systems in Fig. 6 have switched ``cooling flow'' classification in Fig. 7 
(e.g., A1413, A1689).  More importantly is that, {\it in general, the dynamical 
status of a cluster does not seem to influence its position on the $L-T$ 
diagram} (note that the relaxed and unrelaxed NCF clusters lie together in one 
locus while relaxed and unrelaxed CF clusters lie together in another).  This 
point is discussed further in \S 5 and 6.

\begin{deluxetable*}{lccl}
\tablecaption{Properties of nearby clusters based on analysis of
Chandra/XMM-Newton data \label{tab1}}
\tablewidth{35pc}
\tablehead{
\colhead{Cluster} &
\colhead{CF/NCF}  &
\colhead{relaxed?} &
\colhead{references}}
\startdata
2A 0335+096 & CF & yes & Mazzotta et al. (2003)\\
3C 129 & NCF & yes & Krawczynski (2002)\\
A85 & NCF & no & Kempner et al. (2002)\\
A115 & CF & no & Gutierrez \& Krawczynski (2003)\\
A133 & CF & no & Fujita et al. (2002)\\
A478 & CF & yes & Sun et al. (2003a)\\
A496 & CF & yes & Dupke \& White (2003)\\
A644 & CF & no & Lewis \& Buote (2002); Bauer \& Sarazin (2000)\\
A665 & NCF & no & Markevitch \& Vikhlinin (2001)\\
A1060 & NCF & yes & Yamasaki et al. (2002)\\
A1068 & CF & no & Wise et al. (2004)\\
A1367 & NCF & no & Sun \& Murray (2002)\\
A1413 & NCF & yes & Pratt \& Arnaud (2002)\\
A1689 & NCF & yes & Xue \& Wu (2002)\\
A1795 & CF & yes & Ettori et al. (2002)\\
A2029 & CF & yes & Lewis et al. (2003)\\
A2034 & NCF & no & Kempner et al. (2003)\\
A2052 & CF & yes & Blanton et al. (2003)\\
A2142 & CF & no & Markevitch et al. (2000)\\
A2199 & CF & yes & Johnstone et al. (2002)\\
A2218 & NCF & no & Machacek et al. (2002)\\
A2256 & NCF & no & Sun et al. (2002)\\
A2589 & NCF & yes & Buote \& Lewis (2004)\\
A2597 & CF & yes & McNamara et al. (2001)\\
A3112 & CF & yes & Takizawa et al. (2003)\\
A3266 & NCF & no & Henriksen \& Tittley (2002)\\
A3667 & NCF & no & Mazzotta et al. (2002)\\
A3921 & NCF & no & Sauvageot et al. (2001)\\
AWM4 & NCF & yes & O'Sullivan \& Vrtilek (2004)\\
Hydra A & CF & yes & David et al. (2001)\\
MKW4 & CF & yes & O'Sullivan \& Vrtilek (2004)\\
PKS 0745-19 & CF & yes & Chen et al. (2003)\\
RX J1720.0+2638 & CF & no & Mazzotta et al. (2001)
\enddata
\tablecomments{``Cooling flow'' status is based on the
presence or absence of large declining temperature profiles towards the cluster
center while dynamical status is based on the presence or absence of
large-scale
irregularities (presumably due to merging) in the X-ray images of the clusters.
}
\end{deluxetable*}

The $L-M$ relation plotted in Fig. 8 has many of the same features present in the 
$L-T$ relation.  Namely, NCFs and CFs are separated according to luminosity, and 
dynamical status does not seem to systematically affect this trend.  As far as 
we are aware, this is the first time the dispersion in the luminosity-mass 
relation has been shown to depend on ``cooling flow'' status, although Reiprich 
\& B\"{o}hringer (2002) hinted at the existence of such a trend (ApJ, 567, pg. 
730).  Below, we examine whether any of the theoretical models can account for 
this intrinsic scatter and the dichotomy between NCF and CF clusters.

To begin with, we consider the issue of the cooling flow mass deposition 
rate, $\dot{M}$. The new {\it Chandra} and {\it XMM-Newton} high resolution 
X-ray data of massive CF clusters have been used to infer mass deposition rates 
typically ranging from $\dot{M} \sim 100 - 400 M_\odot$ yr$^{-1}$ (e.g., 
Schmidt, Allen, \& Fabian 2001; Ettori et al. 2002; Peterson et al. 2003), 
which is 
substantially lower than previous estimates based on {\it ASCA} and {\it ROSAT} 
data of the same clusters.  Our model predicts mass drop out rates in range $300 
- 500 M_\odot$ yr$^{-1}$ once the initial entropy core has been radiated away.  
To the extent that these two results can be compared, we are comforted by the 
reasonable agreement.  Admittedly, the model results appear to be somewhat 
larger than the ``observed'' values; however, we point out that the latter are 
inferred from the observations assuming an isobaric cooling flow model whereas 
our model values represent the actual physical cooling rate in the systems. 
Strictly speaking, a detailed comparison of the theoretical vs. ``observed'' 
mass 
drop out rates would entail making mock observations of our model clusters and 
then, use the isobaric cooling flow model to infer a mass drop out rate for the 
theoretical models, an exercise that is beyond the scope of the present study.


\subsection{The $L-T$ relation}

Plotted in Figure 9 is a comparison between the theoretical models and the 
observed $L-T$ relation.  The various panels show the predicted relation for a 
range of entropy injection levels.  The thick solid lines in each panel 
represent the luminosity-temperature relation prior to including the effects of 
radiative cooling.  The hatched regions encompass the full range of $L-T$ 
values spanned by the model clusters during 13 Gyrs of cooling (similar to Fig. 
5).  The data is the same as that plotted in Fig. 7.  We have not plotted error 
bars on the data as the statistical error bars on the luminosity are negligible 
and we have restricted ourselves to clusters with temperature determinations to 
better than 20\% (i.e., $\Delta kT_{\rm ew}/kT_{\rm ew} \leq 0.2$).  Thus, the 
intrinsic scatter in the relation dominates the statistical scatter.  

The top left hand panel of Fig. 9 demonstrates why there has been so much 
effort invested in researching the role of non-gravitational gas physics in 
clusters.  The thick solid line in this panel, which represents a model with no 
cooling and no entropy injection, clearly fails to match the observational 
data.  It should be noted that 
{\epsscale{1.2}
\plotone{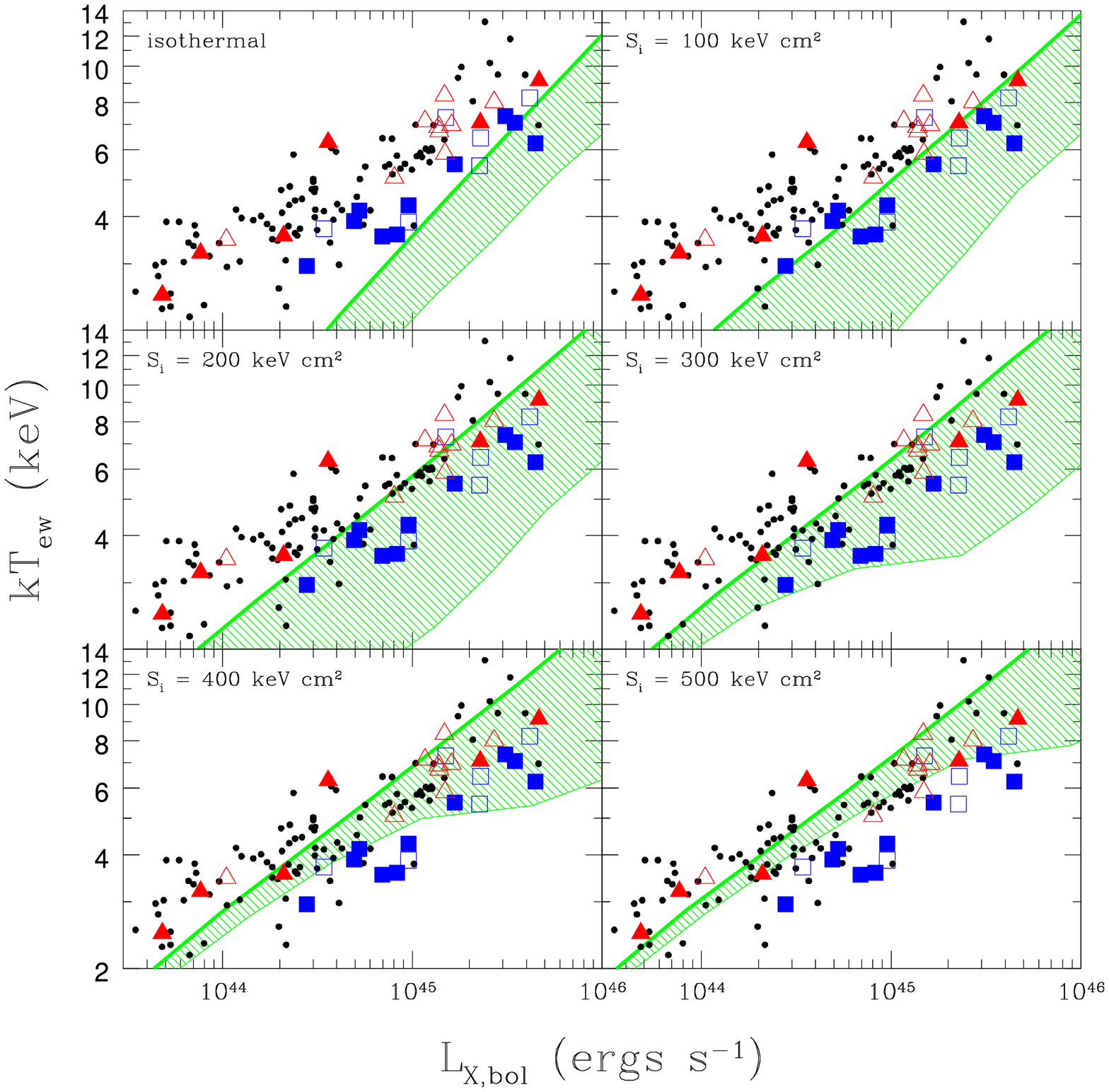}
{Fig. 9. \footnotesize
Comparison of theoretical models to the observed $L-T$ relation.
The thick solid line represent the initial models prior to including the effects
of cooling.  The hatched region represents the full range of predicted $L-T$
values during cooling for 13 Gyr.  The symbols have the same meaning as in Fig.
7.
}}
\vskip0.1in
\noindent
there is essentially no freedom in the isothermal 
model, unless one is willing to resort to a drastically different dark matter 
profile than the one currently being considered.  Cooling the isothermal model 
over the course of 13 Gyr (hatched region) does not improve the situation.  If 
anything, it makes things worse, as the clusters' luminosities increase with 
the inclusion of cooling.  In addition, the model also violates the observed 
cooled gas fraction of clusters reported by Balogh et al. (2001) (see Fig. 1).  
As noted in \S 2.1, very similar results are obtained if, instead of using 
the isothermal model, we make use of the non-radiative self-similar entropy 
profile of Voit et al. (2003) to construct an alternative baseline model.  This 
is compelling evidence that some form of entropy injection is required.

First we investigate models with only a single, fixed entropy injection 
level.  For example, we have previously reported that an entropy injection 
level of $> 300$ keV cm$^2$ 
provides a good fit to the data (e.g., Babul et al. 2002; McCarthy et al. 
2003b).  The aim of these studies was to account for the properties of groups 
and clusters minus any  ``cooling flow'' component and, therefore, the effects 
of radiative cooling were neglected by these models.  A better approach (which 
we have adopted in the present study), however, is to include the effects of 
radiative cooling and explicitly try to model uncorrected data.  Surveying the 
thick solid lines in each of the panels (i.e., entropy injection only models), 
it is seen that injecting the ICM with $\approx 200$ keV cm$^2$ gives the best 
fit to the {\it uncorrected} X-ray data (in the sense that the predicted 
relation falls more or less in the middle of the data).  Injecting $> 300$ keV 
cm$^2$ tends to skew the predicted $L-T$ relation towards lower luminosities 
(i.e., the region occupied by NCF clusters),  confirming our previous results 
for fits to ``cooling flow'' corrected data.  However, with the large amount of 
intrinsic scatter present in the new uncorrected X-ray data, it is clear that no 
single entropy injection level can account for the uncorrected $L-T$ relation.

Alternatively, we could consider a range of entropy injection levels.
As with any physical process, a variation in efficiency is likely to be the 
norm.  However, in order to explain systems on the high-luminosity side of the 
$L-T$ relation (i.e., where CF clusters live), entropy injection levels of $S_i 
\lesssim 300$ keV cm$^2$ are required.  The predicted central cooling time of 
massive clusters with this amount of injection is comparable to 
the Hubble time (see, e.g., Fig. 1) and, therefore, the effects of cooling need 
to be factored in.  Additionally, as pointed out recently by Mushotzky et al. 
(2003), among others, simple entropy injection models cannot account for the 
observed entropy profiles of (some) groups and clusters, most likely because 
of the effects of radiative cooling.  Central positive temperature gradients 
observed in many massive ``cooling flow'' clusters (e.g., Allen, Schmidt, \& 
Fabian 2001) also attest to the importance of radiative cooling.  In other 
words, the observations are unlikely to be explained by a model which includes 
a range of injection levels but not the effects of radiative cooling.

The alternative is to consider a model that includes both entropy injection and 
radiative cooling.  Physically, this is probably the most plausible scenario 
anyway.  However, as demonstrated by Fig. 9, it doesn't seem possible to 
explain all of the data with a single entropy injection level plus radiative 
cooling.  An entropy injection level of $S_i \approx 300$ keV cm$^2$ 
probably comes the closest to explaining the data, but it is clear that a 
significant fraction of the clusters on the low luminosity side of the 
relation are not explained by this model.  Increasing the injection level 
improves the situation but at the expense of losing agreement with the relaxed 
CF clusters.  This makes sense since increasing the injection level mitigates 
the effects of radiative cooling.  We are again forced to consider a model with 
a range of entropy injection levels but this time with the effects of radiative 
cooling included.  With no constraints on the source of the entropy injection, 
indeed it can be seen from the trends in Fig. 9 that the data (including both 
relaxed CF and NCF clusters) can be accounted for by such a model.  In 
particular, the CF clusters can typically be explained by entropy injection 
levels of $S_i \lesssim 300$ keV cm$^2$, while NCF clusters require higher 
injection levels.

\subsection{The $L-M$ relation}

While we place more weight on the $L-T$ relation (because it is based on 
essentially WYSIWYG observables), it is still useful to examine the $L-M$ 
relation as a consistency check.  Figure 10 presents a comparison between 
the observed and predicted luminosity-mass relations$^8$.  Reassuringly, the 
same general trends may be derived from this plot as well; i.e., clusters on the 
low luminosity side of the relation can be explained by high levels of entropy 
injection, while clusters on the high luminosity side can be explained by low 
levels of entropy injection (plus radiative cooling).  There are some 
slight differences, however, in the exact constraints placed on the injection 
levels by the $L-T$ and $L-M$ relations.  These are probably due to issues 
associated with the analysis of the observational data as well as simplifying 
assumptions 
{\epsscale{1.2}
\plotone{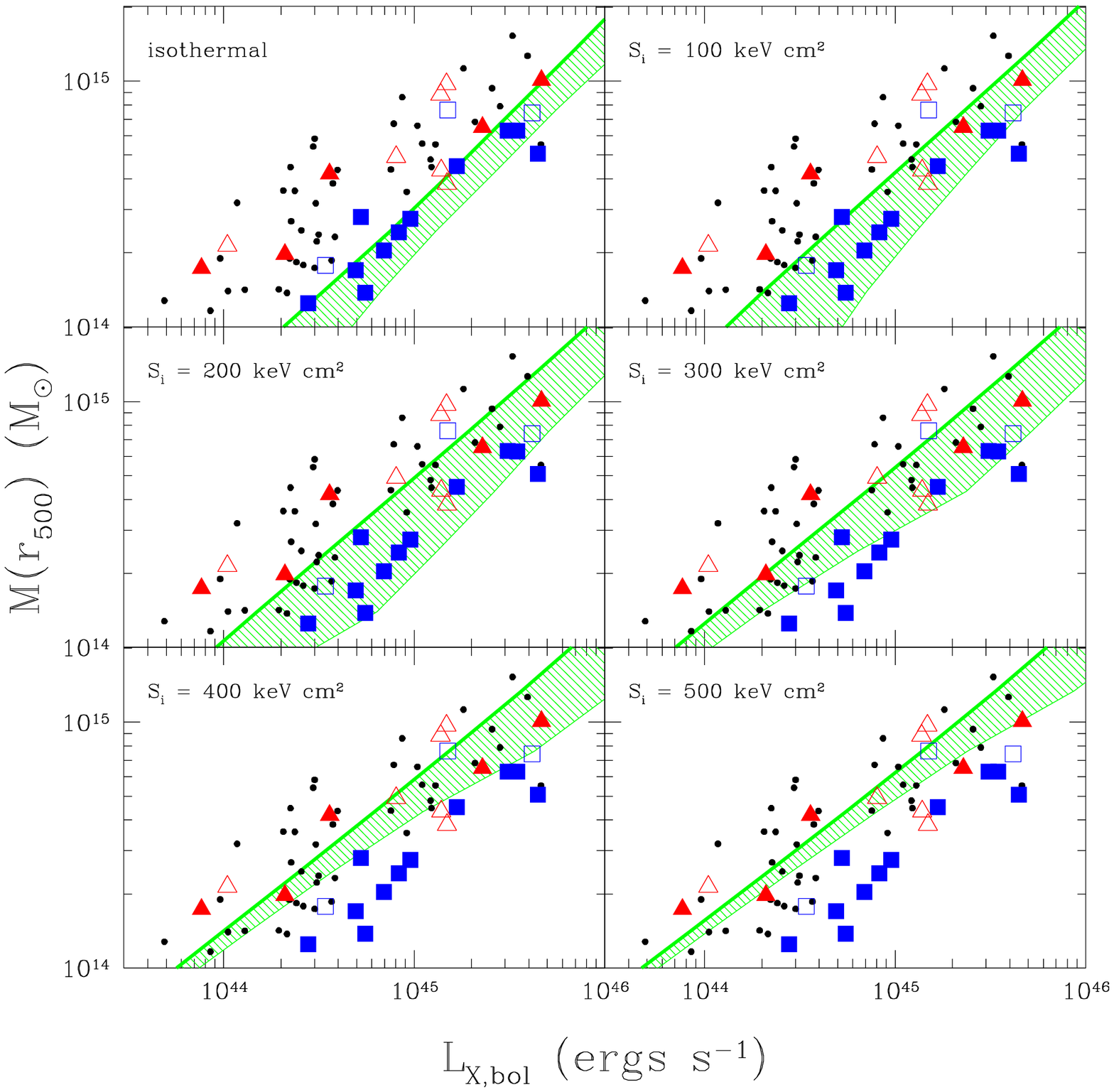}
{Fig. 10. \footnotesize
Comparison of theoretical models to the observed $L-M$ relation.
The thick solid line represent the initial models prior to including the effects
of cooling.  The hatched region represents the full range of predicted $L-M$
values during cooling for 13 Gyr.  The symbols have the same meaning as in Fig.
7.
}}
\vskip0.1in
\noindent
in the modeling [assumptions for the {\it observed mass} calculation 
include isothermality, an assumed form for the surface brightness profiles 
(i.e., beta model), and spherical symmetry].

\footnotetext[8]{The observed values of $M(r_{500})$ refer to the total mass
(baryons and dark matter), whereas the model values neglect the baryon
contribution.  We have ignored the baryonic component in the models as
hydrostatic equilibrium is computed using the dark matter potential only.
However, this is a small effect that we can neglect since, at worst, the model
masses are incorrect by a factor of $\Omega_b/\Omega_m \approx 0.119$ while
the measurements uncertainties are typically $\sim 20$\% (Reiprich \&
B\"{o}hringer
2002).}

\subsection{Summary}

We have found that cooling only and entropy injection only models fail to 
reproduce the observed uncorrected luminosity-temperature and luminosity-mass 
relations of nearby massive, clusters.  However, a model that includes a 
distribution of entropy injection levels and radiative cooling can account for 
these relations, including their intrinsic scatter and the dichotomy between 
relaxed NCF and CF clusters.  In particular, NCF and CF clusters require 
relatively large and small amounts of additional entropy, respectively, with 
$S_i \approx 300$ keV cm$^2$ essentially being the dividing line between the two 
classes of clusters.  In retrospect, this result is not surprising, since 
the cooling threshold, assuming cooling for roughly a Hubble time, for a 
typical massive cluster with $kT_{ew} \sim 6$ keV is $\sim 300$ keV cm$^2$ (see 
Fig. 1 of Voit \& Bryan 2001).

Immediately below, we examine whether or not such a model can account for the 
observed structural properties of clusters, as deduced from new {\it Chandra} 
and {\it XMM-Newton} data.

\vskip 0.2in

{\epsscale{1.2}
\plotone{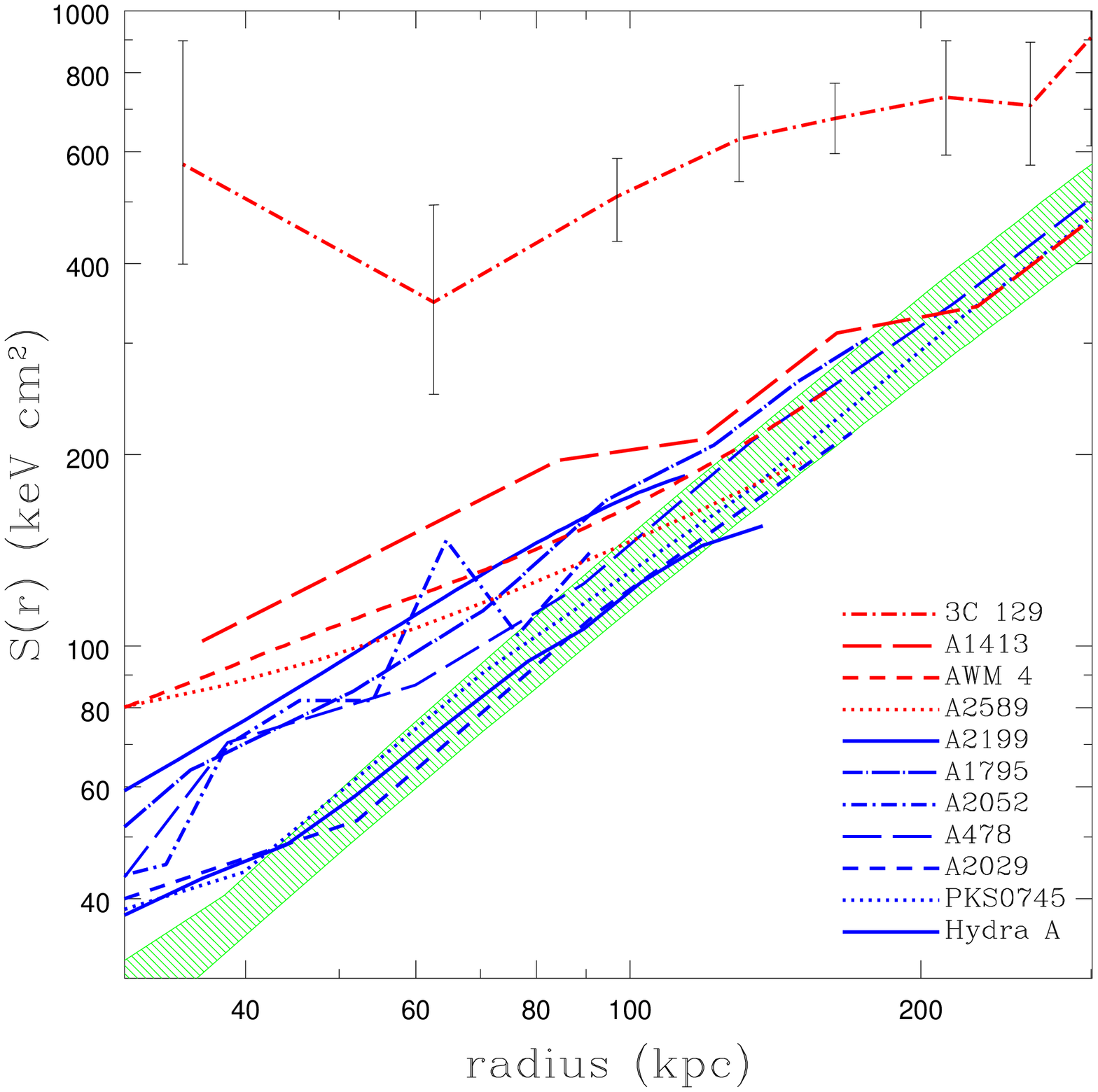}
{Fig. 11. \footnotesize
The entropy profiles of 11 nearby, massive clusters observed with
{\it Chandra} or {\it XMM-Newton}.  The key in the bottom right hand corner is
ordered according to the central entropy values of these clusters in descending
order (i.e., 3C 129 is the top long dashed line, A1413 is the long dashed line
immediately below, and so on).  With the exception of 3C 129, errors on the
derived entropy profiles are typically 10\%.  The hatched region represents the
model predictions for clusters that are actively cooling gas out (see text).
}}
\vskip0.1in

\section{Comparison with Observed Structural Properties}

\subsection{Entropy profiles}

Presented in Figure 11 are the observed entropy profiles of the {\it relaxed} 
clusters labeled in Figs. 7 \& 8 as derived from new {\it Chandra} and {\it 
XMM-Newton} data.  These clusters include 3C 129 (Krawczynski 2002; {\it 
Chandra} data), A1413 (Pratt \& Arnaud 2002; {\it XMM-Newton} data), AWM4
(O'Sullivan \& Vrtilek 2003; {\it XMM-Newton} data), A2589 (Buote \& Lewis 
2004; {\it Chandra} data), A2199 (Johnstone et al. 2002; {\it Chandra} data), 
A1795 
(Ettori et al. 2002; {\it Chandra} data), A2052 (Blanton et al. 2003; {\it 
Chandra} data); A478 (Sun et al. 2003a; {\it Chandra} data), A2029 (Lewis, 
Buote, 
\& Stocke 2003; {\it Chandra} data), PKS0745 (Chen, Ikebe, \& B\"{o}hringer 
2003; {\it XMM-Newton} data) and Hydra A (David et al. 2001; {\it Chandra} data).  
In some cases, the actual deprojected entropy profile was not published but 
could be easily derived from the published deprojected temperature and density 
(or pressure) profiles.  In such cases, we assume a $0.3 Z_{\odot}$ metallicity 
if none was listed.  The hatched region is the predicted zone of entropy 
profiles from clusters with emission-weighted temperatures ranging from $3$ keV 
$\lesssim kT_{\rm ew} \lesssim 9$ keV (roughly matching the observed range) and 
whose {\it initial entropy core has dropped out}.  The value of the initial 
core is irrelevant since, once the core has dropped out, the resulting entropy 
profiles for clusters of a given mass are identical in the model of Babul et 
al. (2002).  In other words, the hatched region illustrates where clusters that 
are actively cooling gas out should live.  If, however, the gas was injected 
with entropy and has not had enough time since to cool out the resulting 
entropy core, we should expect to see elevated entropy levels near the cluster 
center (see Fig. 2).  At large radii, however, all of  the profiles should 
converge to the hatched region, as this is the regime where shock heating 
becomes much more important than the non-gravitational entropy injection. 

One of the first features that leaps out of Fig. 11 is that essentially all of 
the observed entropy profiles converge to the hatched region at large radii, 
where non-gravitational physics is less important.  This suggests that 
high resolution numerical simulations are doing an excellent job of capturing 
the gravitational gas physics of cluster formation (again, we note that the 
entropy profile at large radii in our models has been forced to match the 
results of non-radiative simulations).  

But how does the entropy injection plus cooling model measure up to the 
observed entropy profiles at small radii?  The answer is surprisingly well.  A 
comparison of Fig. 11 to Figs. 7 \& 8 shows that there is a clear trend between 
central entropy and dispersion in the $L-T$ and $L-M$ relations.  Namely, the 
central entropy increases as one goes from the high luminosity side to the the 
low luminosity side of these relations.  This is quite reminiscent of 
the trend between dispersion in the $L-T$ and $L-M$ relations and ``cooling 
flow'' status plotted in Figs. 6-8.  In fact, based on the present results, we 
would argue that the latter is a direct consequence of the former.  For example, 
in order to explain an `extreme' (relaxed) NCF cluster like 3C 129 on the basis 
of the results presented in \S 4, we require an extraordinarily large entropy 
injection level of $S_i > 500$ keV cm$^2$ (see Fig. 9).  This may seem 
unlikely, but Fig. 11 indeed demonstrates that 3C 129 has a very high central 
entropy.  Also consistent with this trend are intermediate clusters such as 
A2589, AWM 4, and A1413 (which lie more or less in the middle of the $L-T$ and 
$L-M$ relations), that show elevated central entropies, and massive ``cooling 
flow'' clusters, such as PKS0745 and Hydra A, which show almost no excess 
entropy in their cores.

So the entropy injection plus cooling model we have proposed works fairly
well in terms of explaining the observed entropy profiles of massive clusters.  
However, the current sample of 11 profiles is obviously too small to be 
definitive.  It should also be noted that the majority of clusters with 
published entropy profiles were selected on the basis of prior 
(i.e., {\it ROSAT}/{\it ASCA}) evidence for ``cooling flows'' (see Figs. 7 \& 
8).  Therefore, it is reasonable to expect that the current data set is 
biased towards clusters with low central entropies.  This may be partially 
responsible for earlier claims that large amounts of entropy injection are 
ruled out (e.g., Pratt \& Arnaud 2003; Mushotzky et al. 2003).  A simple way of 
testing this model would be to obtain the entropy profiles for 
a large, {\it representative} sample of clusters.  It should be kept in mind 
that NCF clusters could make up 30\% (or more) of all nearby clusters (Peres et 
al. 1998).  We expect that such samples will soon be available as more and more 
{\it Chandra} and {\it XMM-Newton} data become public.

{\epsscale{1.2}
\plotone{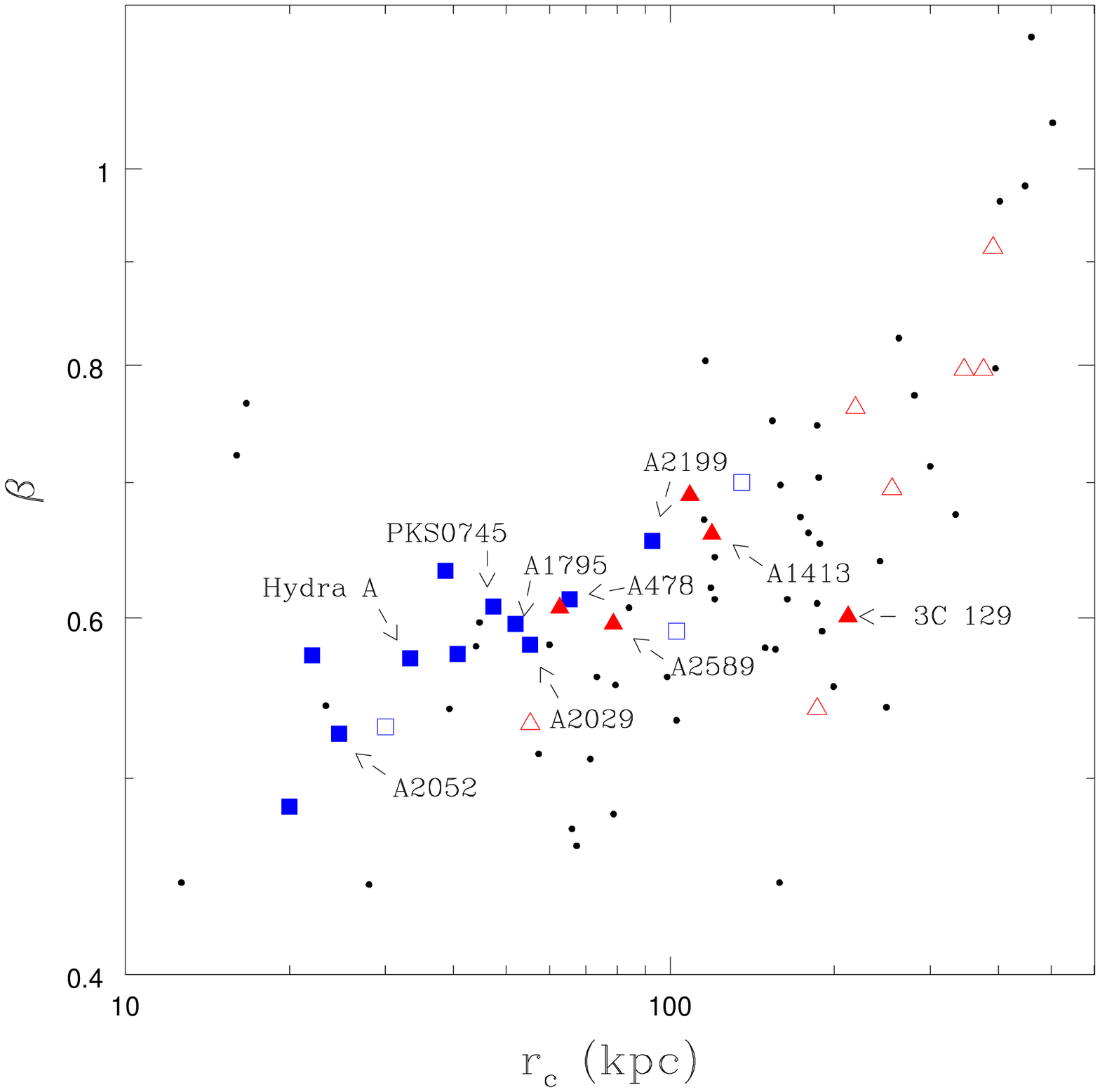}
{Fig. 12. \footnotesize
The $\beta - r_c$ relation for nearby, massive clusters observed
with {\it ROSAT} by Reiprich \& B\"{o}hringer (2002).  The symbols have the same
meaning as in Fig. 7.  Labeled clusters are those with published entropy
profiles (see Fig. 11).
}}
\vskip0.1in

\subsection{Surface brightness profiles}

X-ray observers often parameterize the observed X-ray surface brightness
profiles by fitting the observed profile to assumed analytic functions.
The simplest and most commonly used of these is the $\beta$ model (Cavaliere 
\& Fusco-Femiano 1976).  This family of curves is characterized by two 
parameters: $r_c$, which measures the size of the constant surface brightness 
X-ray core, and $\beta$, which measures how rapidly the X-ray flux falls off 
with radius beyond the core.  While we recognize that the simple $\beta$ model
does not provide a perfect match to {\it either} the observed profiles
or the predicted profiles of our model clusters, the approach does
provide a useful characterization of the data/models for the purposes
of comparison especially since the values of $\beta$ and $r_c$ for the
observed profiles are readily available in literature.

Plotted in Fig. 12 is the observed relationship between the surface 
brightness shape parameters, $\beta$ and $r_c$, as determined by Reiprich \& 
B\"{o}hringer (2002) by fitting the single $\beta$ model to clusters in the 
extended {\it ROSAT} HIFLUGCS sample.  Only those clusters in common with the 
ACC have been plotted.  For clarity, we have not plotted measurement error bars.  
Typically, measurement uncertainty on $\beta$ and $r_c$ is $\lesssim 10$\%.

One immediately noticeable trend is that the ``cooling flow'' and ``non-cooling 
flow'' clusters are separated according to core radius size and, to a 
much lesser extent, by the value of $\beta$.  This trend between core radius 
size and ``cooling flow'' status has been known for some time (e.g., Mohr, 
Mathiesen, \& Evrard 1999; Ota \& Mitsuda 2002).  More interestingly, the 
unrelaxed systems (particularly the unrelaxed NCF clusters) appear to be 
separated from the relaxed systems, in the sense that systems that are 
undergoing (or have recently undergone) mergers have larger core radii than 
those systems which 
{\epsscale{1.2}
\plotone{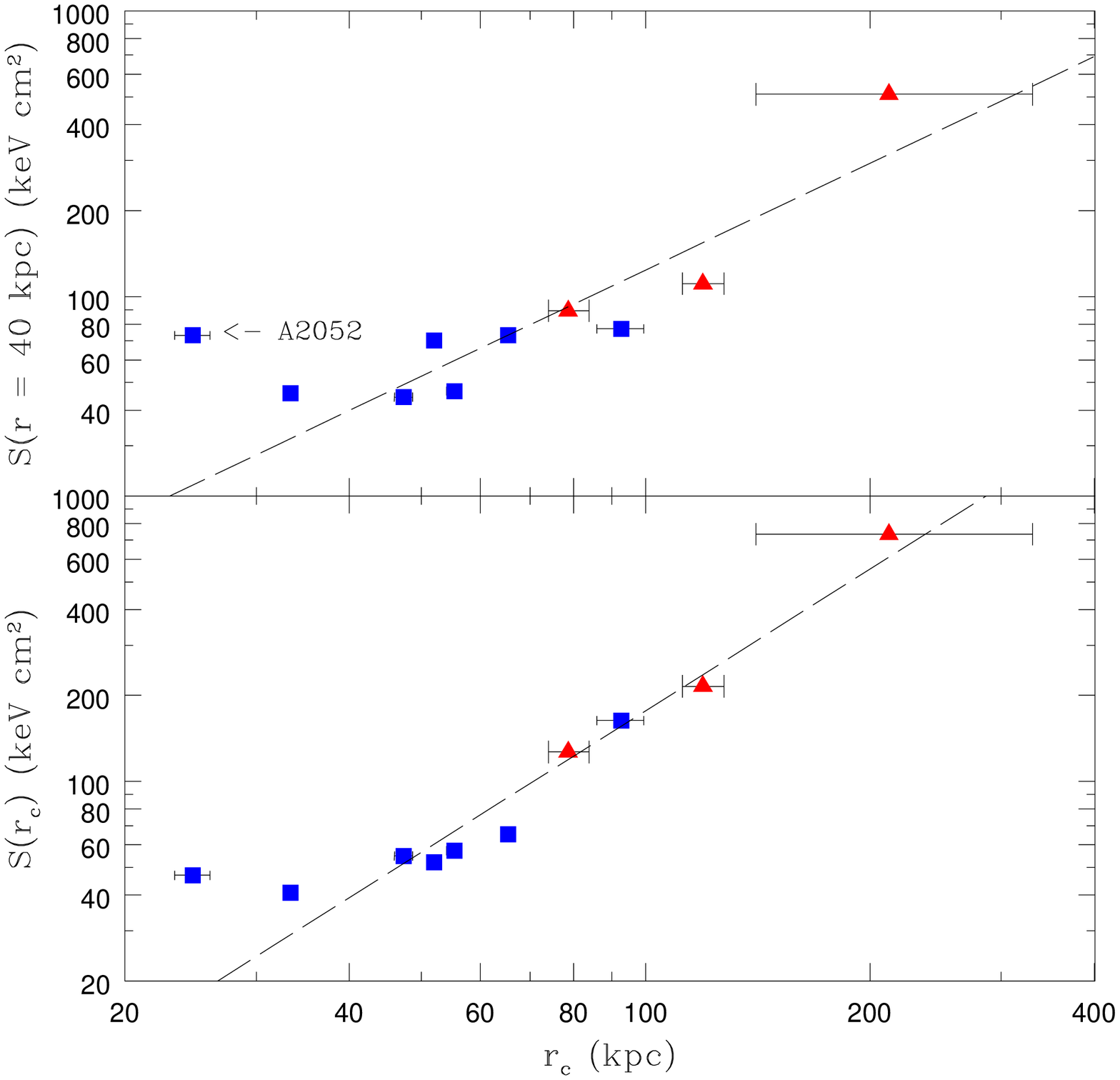}
{Fig. 13. \footnotesize
The observed relationship between core radius size and central
entropy for the clusters plotted in Fig. 11.  {\it Top}: Entropy at 40 kpc.
{\it Bottom}: Entropy at the core radius.  Measurement uncertainty in the
entropy is typically 10\%.  Symbols have the same meaning as in Fig. 7.  Dashed
lines indicate the best fit power-laws to the relations (excluding A2052, see
text).  Namely, $\log{[S(r=40 {\rm kpc})]} = 1.24 \log{r_c} - 0.39$ and
$\log{[S(r_c)]} = 1.65 \log{r_c} - 1.05$.
}}
\vskip0.1in
\noindent
appear relaxed.  Given the results of \S 4.1, therefore, 
it would seem that mergers tend to influence the structural profiles of 
clusters (within the central regions) but not the overall global properties.  
This is consistent with the fact that ``cooling flow correction'' of relaxed 
CF clusters results in global properties typical of (relaxed and unrelaxed) NCF 
clusters (e.g., the ``cooling flow corrected'' $L-T$ relation; Markevitch 1998).

In \S 5.1, we found a correlation between dispersion in the $L-T$ and $L-M$ 
relations and the central entropy of clusters.  Comparison of Fig. 12 with 
Figs. 7 and 8 illustrates that there is also a correlation between dispersion 
in the scaling relations and the size of a cluster's core radius.  Therefore, 
we should expect an observed relationship between the size of a cluster's core 
radius and its central entropy.  These quantities are plotted in Fig. 13 for the 
{\it relaxed} clusters explicitly labeled in Figs. 11 and 12.  In the top panel 
of Fig. 13, we plot the entropy at a radius of 40 kpc versus core radius size, 
while in the bottom panel we plot the entropy at the core radius versus the 
core radius size.

With the exception of A2052, the results plotted in the top panel of Fig. 13 
clearly demonstrate that the core radius of a cluster increases with increasing 
central entropy.  The dashed lines show the best fit power-law relationship 
between the central entropy and core radius excluding A2052.  The bottom panel, 
which instead plots the entropy at the core radius versus core radius size, 
shows an even tighter relationship with $\log{S(r_c)} = 1.65 \log{r_c} - 1.05$ 
(with $r_c$ in kpc and $S(r_c)$ in keV cm$^2$).

A possible explanation for why A2052 is scattered away from the trend
traced out by the rest of the clusters in 
{\epsscale{1.2}
\plotone{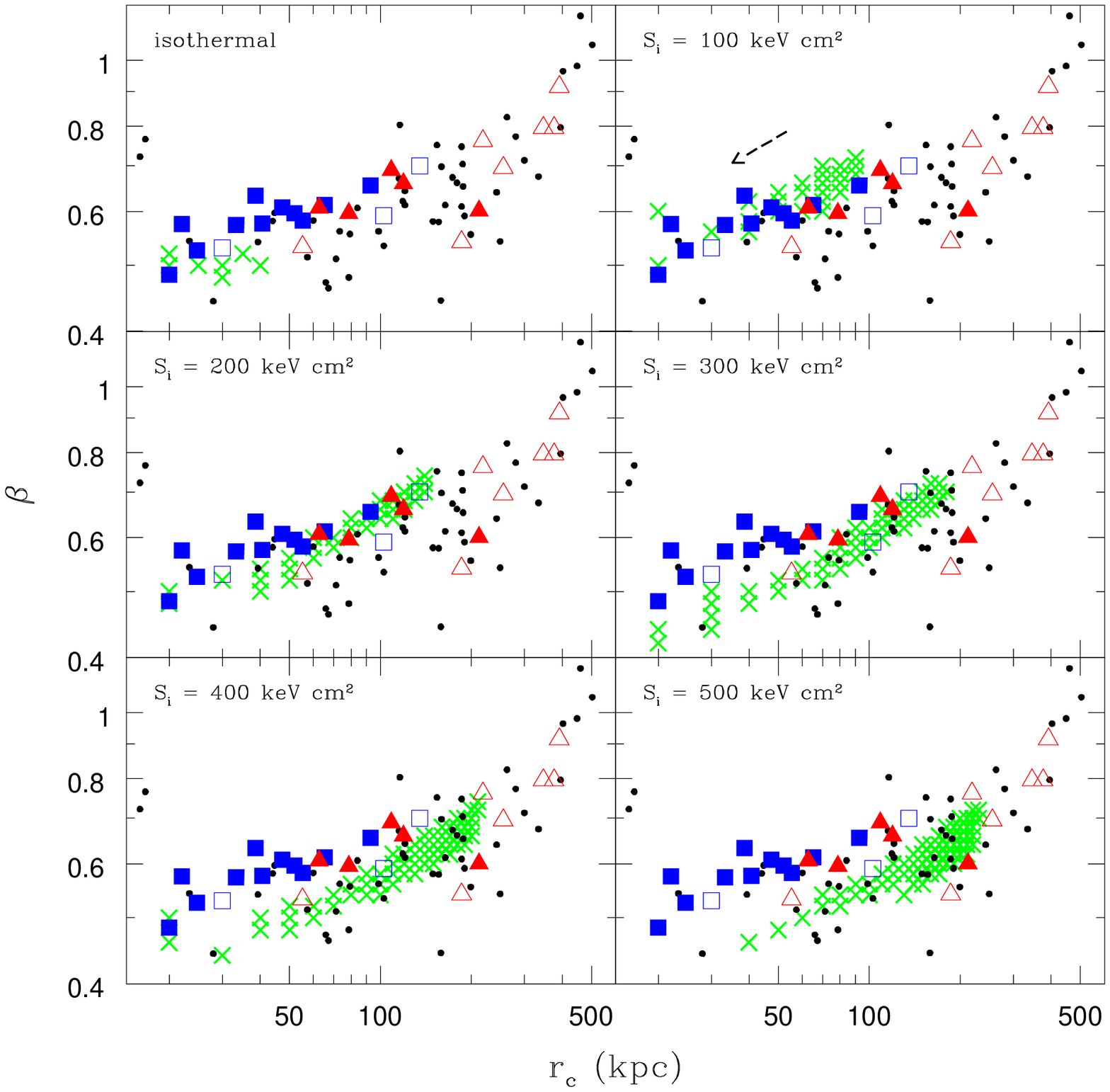}
{Fig. 14. \footnotesize
Comparison of theoretical models to the observed $\beta-r_c$
relation.  The symbols have the same meaning as in Fig. 7 with the exception of
the diagonal crosses, which represent the model predictions (see text).  The
arrow in the upper right hand panel indicates which way the model predictions
evolve as the clusters cool.
}}
\vskip0.1in
\noindent
Figure 13 is the presence of
prominent bubbles in the core of this cluster (Blanton et al. 2003).  While it 
is true that a number of other CF clusters plotted in Figure 13 also show 
evidence for bubbles in central regions (eg. A2199 and Hydra A), the bubbles in 
these clusters are relatively small compared to size of the region over which 
the gas is cooling (i.e., region within which the gas temperature declines 
towards the cluster center).  A2052, by contrast, has ``large'' bubbles and the 
impact 
of these features is such that one can account for the ``cooling flow'' 
properties of A2052 by azimuthally averaging over the X-ray bright, cool, bubble
shells (McCarthy et al. 2003c).   We hypothesize that the anomalous values
of $\beta$ and $r_c$ for A2052 are the result of contamination from these
bubbles.   This illustrates a potential pitfall of characterizing the surface 
brightness profiles of clusters with simple $\beta$ models, in that substructure 
can potentially throw off the fit to the data and give misleading results.  
Nonetheless, we find it quite remarkable that, within the context of the small 
sample we have examined, most systems follow a tight trend between core radius 
size and central entropy.

It is interesting to see whether or not the models can explain the $\beta 
- r_c$ relation plotted in Fig. 12 and also account for the relationship between 
the central entropy and $r_c$ as well.  Plotted in Fig. 14 is a comparison of the 
observed and predicted $\beta - r_c$ relations.  The symbols have the same 
meaning as in Fig. 7 with exception of the diagonal crosses, which represent the 
model predictions.  The crosses are the result of fitting single $\beta$ models 
to the model surface brightness profiles as a function of time (as the cluster 
cools) for a suitable range of cluster masses [i.e., $M(r_{500}) \gtrsim 10^{14} 
M_{\odot}$].  As the clusters cool, both $\beta$ and (especially) $r_c$ decrease 
(i.e., with time, the model predictions typically move from upper right hand side 
of the plot to the lower left hand side; see arrow in upper right hand panel).

The results of Fig. 14 illustrate that models with entropy injection levels of 
$S_i \lesssim 200$ keV cm$^2$ + radiative cooling can account for relaxed CF 
clusters, while higher levels of entropy injection are typically required to 
explain relaxed NCF clusters.  This is in excellent agreement with the results 
of \S 5.  Not surprisingly, the unrelaxed NCF systems have core radii that are 
much larger than can be accounted for by the model clusters (which, by 
definition, are relaxed).  Therefore, core radius size is a potentially promising 
way of distinguishing between relaxed and unrelaxed systems.  In addition, the 
crosses in Fig. 14 indicate that as one injects more and more entropy into 
a system the larger its core becomes (as is also evident from Figs. 2 and 3).  
This qualitatively matches the trends seen in Fig. 13.  Because the central 
entropy and core radius evolve with time in the models, a more direct comparison 
between theory and observations would require knowledge about how long each 
observed system has been able to cool for.

\subsection{Temperature profiles}

Finally, we make a comparison between observed and predicted temperature 
profiles.  Plotted in Fig. 15 is the so-called universal temperature profile 
of Allen, Schmidt, \& Fabian (2001).  This was derived from a sample of 6 
clusters observed with {\it Chandra} and classified as relaxed, massive 
``cooling flow'' clusters by these authors.  Also shown (solid lines) are 
the predictions of our entropy injection plus cooling model {\it after the 
initial core has dropped out} for a range of cluster masses.  The model 
profiles have been normalized the same way as the observational results; i.e., 
radii have been normalized to $r_{2500}$ and temperatures have been normalized 
to the temperature at that radius.

Overall, the fit to observational data is quite good.  There is a hint of 
some slight differences at very small radii, although the hatched region 
(which represents Allen, Schmidt, \& Fabian's best fit to their data) does not 
encompass all of the scatter in the observed temperature profiles (see Fig. 1 
of Allen, Schmidt, \& Fabian 2001).  Thus, so long as there is enough time to 
cool out the initial entropy core, the model does a very good job of matching 
relaxed, massive CF clusters.  

Unfortunately, there are very few relaxed NCF clusters with published {\it 
Chandra} or {\it XMM-Newton} temperature profiles.  Furthermore, comparison with 
observations is made difficult by the fact that our model does not predict a 
quasi-steady state ``non-cooling flow'' temperature profile but, rather, a 
full distribution of profiles (see Fig. 4).  We do note, however, that 3C 129 has 
a sharp {\it negative} temperature gradient at its center (i.e., sharp rise 
towards the center; see Fig. 2 of Krawczynski 2002), which is qualitatively 
what we should expect from a system that had a large injection of 
entropy (see Fig. 4).

\section{Discussion}

A major result of the present study is that a relatively wide range in 
entropy injection levels combined with subsequent radiative cooling is 
required to explain the observed global and structural properties of massive 
clusters.  That cooling is an essential process is the least surprising 
result.  The very fact that clusters are radiating in X-rays is evidence that 
they are cooling.  However, the large scatter in their global and structural 
properties suggests large cluster-to-cluster variations in cooling efficiency.  
We have shown that this can be accounted for as a by-product of variations in 
the level of entropy injection.  Regardless of the source of the entropy 
injection, cluster-to-cluster variations in injection efficiency are to be 
expected.  A similar suggestion was recently put forward by Sun et al. (2003b) 
in order to explain scatter in the observed entropy profiles of 6 low mass 
groups.  

\footnotetext[9]{Since only a small fraction of the clusters examined in
the present study have high quality {\it Chandra} or {\it XMM-Newton} images
available, we have not explicitly excluded merger systems from the first part
of
this study.  As we have shown, mergers do not seem to significantly influence
{\it global} properties such as the $L-T$ and $L-M$ relations (see \S4.1) and,
therefore, we do not expect a significant bias to be present.  When
investigating {\it structural} properties (\S 5), on the other hand, only the
profiles of relaxed systems were considered, as azimuthally-averaged profiles
for highly asymmetric (merging) clusters are essentially meaningless.}

Variations in the injection level can also naturally account for the dichotomy 
between NCF and CF clusters.  The prevailing view is that NCF clusters are 
systems that have been disrupted by recent major mergers.  This picture is 
often supported by images of NCF clusters which show disturbed X-ray 
morphologies (although, admittedly, NCF clusters typically have flatter surface 
brightness profiles and, therefore, it should be easier to pick out 
irregularities in these systems than in CF clusters).  Indeed, it is likely 
that some of the NCFs are the result of mergers, and such systems should be 
removed from consideration$^9$, but, as we have already discussed, there 
{\epsscale{1.2}
\plotone{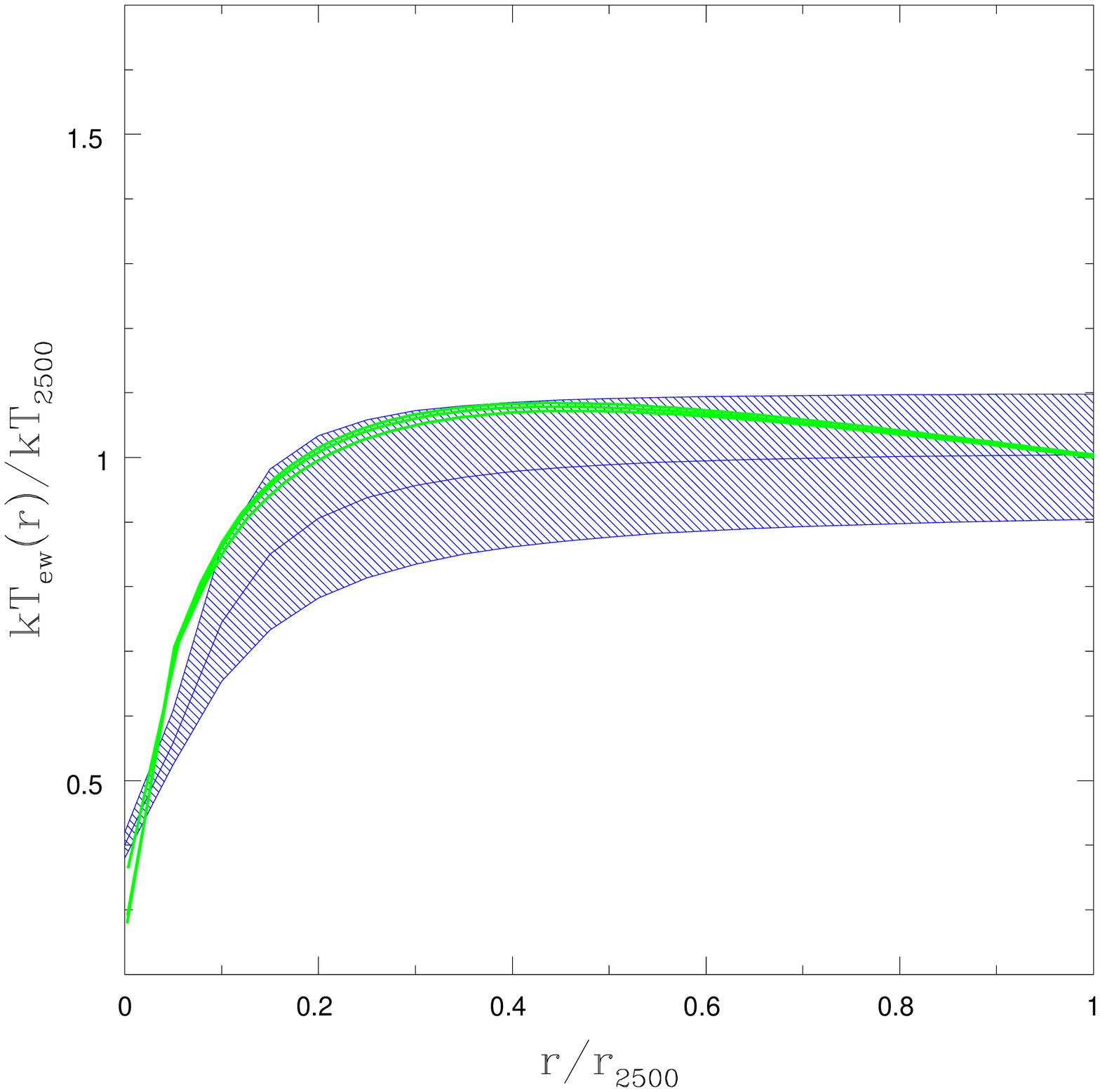}
{Fig. 15. \footnotesize
Comparison of observed and predicted temperature profiles for
``cooling flow'' clusters.  The hatched region represents the observational
results of Allen, Schmidt, \& Fabian (2001) from a sample of 6 massive,
``cooling flow'' clusters.  The solid black lines represent the predictions of
our model {\it after the entropy core has dropped out} for a range of high mass
clusters.  
}}
\vskip0.1in
\noindent
are several NCF clusters that look remarkably relaxed and do not show any 
obvious signs of ongoing mergers 
(e.g., 3C 129, A2589, A1060, A1651, A1689, 
AWM 4, A3571, A1413, RX J1200.8-0328, RX J1120.1+4318).  Such systems need to 
be accounted for.  
Moreover, in the currently favored hierarchical model for structure formation, 
{\it all massive systems} are formed through mergers and accretion of smaller 
objects.  Therefore, CF clusters should also experience major mergers, since 
they are a ubiquitous feature of the $\Lambda$CDM cosmology.  In fact, Loken, 
Melott, \& Miller (1999) showed that CF clusters tend to occupy more crowded 
regions of the universe than NCF clusters.  Perseus, RX J1347.5-1145, A2142, 
A1644, and A644 are all examples of clusters which are believed to be undergoing 
mergers (or have recently undergone a merger) and yet have retained their 
``cooling flow'' identity.  This argues against merging being the only difference 
between CF and NCF clusters. 

In order to explore these ideas more explicitly, we have carried out a series 
of numerical simulation experiments involving the merging of clusters of 
various mass ratios and impact parameters.  In addition, the effects of 
radiative cooling have been included.  The results of this study will be 
presented in a forthcoming paper (Poole et al. in preparation).  We briefly 
note, however, that the recent hydrodynamic study of Motl et al. (2004) shows 
that it is extremely difficult to disrupt the dense cores of cooling clusters 
and, subsequently, form a system that resembles a NCF cluster.  Typically, the 
clusters regain their CF status on a very short time scale.  Similar 
conclusions were reached by G\'{o}mez et al. (2002).  This lends further 
credence to the hypothesis that some other mechanism, in addition to merging, 
may be required in order to explain the relatively large fraction of NCF 
clusters observed in the local universe.  Based on the results of the current 
study, we would argue entropy injection is the required mechanism.

But what is the source of the additional entropy?  This is perhaps the 
single most 
important outstanding issue of models which implement entropy injection.  A 
whole host of entropy injection mechanisms (a number of which are discussed in 
detail in \S 1 of Babul et al. 2002) have been proposed to explain the 
overcooling/cooling flow problem and also the global X-ray scaling relations of 
clusters.  The most commonly proposed mechanism is galactic winds 
driven by supernovae, which are expected to transfer relatively large amounts 
of thermal energy into the ICM (e.g., Loewenstein 2000; Voit \& Bryan 2001).  
However, in order to explain the scaling relations of clusters, the efficiency 
of the supernovae must be extremely high (probably unreasonably high) and even 
then one requires a top-heavy stellar initial mass function (e.g., Balogh et 
al. 1999; Valageas \& Silk 1999).  As such, supernovae by themselves are 
unlikely to be the source of the entropy/energy injection.  Heating via quasars 
and active galactic nuclei (AGN), either prior to or following cluster 
formation, is a currently popular (proposed) mechanism, since such objects 
potentially contain vast reservoirs of energy.  While it seems there is plenty 
of energy available, it is still unclear exactly how the AGN heat the gas.  
Possibilities include Compton heating of intracluster electrons via high energy 
UV and X-ray photons emitted by the AGN accretion disk (e.g., Ciotti \& 
Ostriker 1997, 2001), shock heating by transonic or supersonic jets (e.g., 
Binney \& Tabor 1995; Omma et al. 2004), entrainment, transport, and 
subsequent mixing of low entropy gas via buoyantly rising bubbles of hot plasma 
inflated by the AGN (e.g., Quilis, Bower, \& Balogh 2001; Mathews et al. 2003; 
Dalla Vecchia et al. 2004), and viscous dissipation of the kinetic energy of the 
rising bubbles (e.g., Fabian et al. 2003; Ruszkowski, Bruggen, \& Begelman 
2004).  Detailed calculations, which are beyond the scope of this paper, are 
required in order to assess whether AGN are able to give rise to the 
distribution of entropy injection levels required to account for the X-ray 
properties discussed above.

Entropy injection into the central regions of the clusters can also be
achieved via heat transport from outer regions.   Two transport mechanisms have 
been proposed: thermal conduction (e.g. Narayan \& Medvedev 2001) and turbulent 
mixing (e.g. Kim \& Narayan 2003).  Until recently it was thought that 
conduction would be an inefficient heat transport mechanism, since the presence 
of intracluster magnetic fields should strongly suppress conduction.  However, 
Narayan \& Medvedev (2001) demonstrated that if the 
magnetic fields are tangled it is still possible to achieve conductivities of 
up to one-third the Spitzer conductivity.  Zakamska \& Narayan (2003) 
subsequently demonstrated with simple models that conduction could offset 
radiative losses in some (but not all) CF clusters.  However, through the use 
of hydrodynamical simulations that include radiative cooling and conduction, 
Dolag et al. (2004) showed that while conduction may be important for the most 
massive clusters, it does not significantly modify the properties of lower 
mass clusters (since the conductivity has a strong temperature dependence).  
Thus, thermal conduction is unlikely to be solely responsible for the 
distribution of entropy injection levels inferred in the present study.  On the 
other hand, turbulent mixing of the intracluster gas, which is likely caused by 
stirring due to infalling and orbiting substructure, is perhaps more promising 
(El-Zant et al. 2004; Bildfell et al. in preparation).  For example, Kim \& 
Narayan (2003) have demonstrated that one can reproduce the slope of the 
$L-T$ relation for high mass clusters with a simple mixing model.  However, it 
has yet to be demonstrated whether such a model can account for the 
normalization of the relation and its associated scatter.

Finally, we note that our estimates for the amount of non-gravitational entropy 
required to explain the observed x-ray properties of relaxed clusters --- and 
our subsequent evaluation of the possible sources of this entropy --- implicitly 
assumed that in the absence of any entropy injection, the cluster gas will 
exhibit a very small entropy core, if any.  The reason for making this 
assumption is because if one places isothermal gas in the cluster potential well
(i.e., the standard model), the gas density distribution will have an 
exponential form and, correspondingly, the entropy profile will have a small 
core (both in size and amplitude).  This implicitly assumes the underlying 
(dark matter) potential is similar to those found from high resolution 
numerical simulations.  For a singular isothermal sphere, there is no entropy 
core whatsoever.

The above analytic result is borne out by SPH non-radiative simulations of
galaxy clusters.  Although there is some evidence for an entropy core in low
resolution simulations,  this core decreases in size and amplitude with
increasing resolution (e.g., Frenk et al. 1999; Lewis et al. 2000), indicating 
that the core in the low-resolution simulations is a numerical artifact.

More recently, though, high-resolution mesh-based (e.g., AMR) non-radiative 
simulations show substantial entropy cores that persist even when the simulation 
resolution is increased, while agreeing with SPH results at large radii (e.g., 
Voit et al. 2003).  Moreover, there appears to be a coupling between the 
magnitude and the size of the entropy core in individual cluster and the 
cluster's merger history.  The origin of the core, and in fact whether
the core is indeed real has yet to be ascertained.  In fact, there is no clear 
explanation available as to why the SPH and mesh-based codes give such different 
result.  Clearly, though, a more careful study is warranted because if the AMR 
results are correct, then the stringent constraints on the amount of 
non-gravitational entropy injection required is correspondingly reduced.  Quite 
independent of how the entropy core arises, our present study demonstrates that 
a distribution in magnitudes of the central entropy cores in galaxy clusters is 
key to understanding their global and structural properties.

Finally, in the present study, we have purposely focused on high mass clusters 
since temperatures and luminosities are fairly straightforward quantities to 
measure for these systems.  However, a complete picture must also 
address the intragroup medium.  The low X-ray luminosity of poor groups makes 
this a difficult challenge.  Not only are there much poorer statistics, in terms 
of low signal-to-noise ratios, but a significant fraction of the flux can
originate from point sources and the ISM of the central galaxy, making
disentanglement of the ICM that much harder.  A good example of just how
difficult it is to obtain reliable results from group data is presented
in the recent study of Osmond \& Ponman (2004).  By simply increasing the
number of low temperature systems in their data set, and also correcting the
luminosity to a fixed overdensity, these authors found that the evidence
for steepening of the $L-T$ relation in groups (e.g., Mulchaey \&
Zabludoff 1998; Helsdon \& Ponman 2000), relative to self-similar predictions,
is no longer solid.  In order to make a fair comparison
between the models and group data, one needs to fold the instrumental response
of the X-ray satellite into the theoretical models.  Furthermore, the models
should be analysed the same way as the observational data (Poole et al.
2004b in preparation).  A future project, therefore, is to
`observe', with a mock X-ray satellite (which mimics {\it Chandra} or {\it
XMM-Newton}), a realistic population of groups and clusters that includes the 
effects of radiative cooling and specific entropy injection processes such as 
conduction, mixing, and AGN heating.

\section{Conclusions}

Recent X-ray observations have highlighted the lack of large isentropic cores
in groups and clusters and have led some to suggest that radiative cooling
is the dominant mechanism in the breaking of self-similarity.  However,  
radiative cooling alone leads to a predicted overabundance of cooled gas in 
theoretical models (the so-called ``cooling crisis'').  In order to 
explain the observed entropy profiles and also retain consistency with the 
observed fraction of cold baryons (stars) in clusters, it is likely
that both radiative cooling and some form of entropy injection are required.
We have performed a thorough investigation of this scenario by adding a
realistic treatment of the effects of radiative cooling to the entropy
injection model of Babul et al. (2002).  A comparison to the current suite of 
X-ray observations was then made with a particular emphasis on assessing 
whether or not the model could account for the large amount of intrinsic 
scatter in the observed scaling relations and, simultaneously, account for the 
observed entropy, surface brightness, and temperature profiles of clusters.  The 
main results can be summarized as follows:

\begin{itemize}

\item{Injecting the ICM with $S_i \gtrsim 200$ keV cm$^2$ prevents significant 
mass drop out due to radiative cooling and insures the observed fraction of cold 
baryons ($\lesssim 10$\%; Balogh et al. 2001) is not violated.}

\item{Radiative cooling approximately maintains the initial power-law 
between entropy and radius, $S \propto r^{1.1}$, at large radii and extends it 
to small radii as well, so long as there is enough time for cooling to wash out 
the effects of any non-gravitational entropy injection.}

\item{Depending on the time elapsed since entropy injection (i.e., the time 
available for the cluster to cool radiatively), the model naturally predicts 
either clusters with flat central surface brightnesses and sharp central  
negative temperature gradients or clusters with peaked surface brightnesses and 
central positive temperature gradients.  These match the main qualitative 
features of NCF and CF clusters, respectively.}

\item{Radiative cooling tends to have a larger effect on the luminosity of a 
cluster than its temperature.  So long as only a small fraction of the 
cluster's baryons are cooled out ($\lesssim 10$\%), the result is that cooling 
moves the predicted $L-T$ relation to higher luminosities at a fixed 
temperature.}

\item{An analysis of the $L-T$ and $L-M$ relations derived from the ACC of 
Horner (2001) and the extended HIFLUGCS of Reiprich \& B\"{o}hringer (2002) 
demonstrates that there is a strong correlation between dispersion in these 
relations and ``cooling flow'' status.  Although this trend has been 
illustrated before for the $L-T$ relation (although probably not as clearly), 
this is the first time it has been demonstrated for the $L-M$ relation.}

\item{We find that a distribution of entropy injection levels combined with the 
effects of radiative cooling can account for the observed $L-T$ and $L-M$ 
relations and the large amount of intrinsic scatter associated with each.  This 
is the first time a theoretical model has been shown to account for this 
intrinsic scatter.  We also find that so-called ``cooling flow'' clusters 
typically require `mild' amounts of entropy injection ($S_i \lesssim 300$ keV 
cm$^2$), whereas ``non-cooling flow'' clusters require larger amounts of 
injection.  Interestingly, this dividing line is  essentially equal to the 
cooling threshold for massive clusters (Voit \& Bryan 2001; Babul et al. 
2002), implying that the amount of entropy injection dictates which class (CF or 
NCF) a particular cluster will fall under.  Moreover, so long as the CF 
clusters were injected with $S_i \gtrsim 200$ keV cm$^2$, their 
predicted cold gas fractions do not violate observational constraints.}

\item{A natural consequence of our explanation for the scatter of the $L-T$ 
and $L-M$ relations is that the central entropy of clusters should increase as 
one goes from the high-luminosity side of these scaling relations to the 
low-luminosity side.  An examination of the deprojected entropy profiles of 11 
relaxed massive CF and NCF systems observed with {\it Chandra} and/or {\it 
XMM-Newton} reveals just such a correlation between central entropy and 
dispersion in these scaling relations.  It is this trend which likely gives rise 
to the previously identified relationship between $L-T$ dispersion and inferred 
cooling flow strength (see Fabian et al. 1994).}

\item{The model predicts surface brightness profiles that are consistent 
with those of relaxed CF clusters when mild amounts of entropy are injected 
($S_i \lesssim 200$ keV cm$^2$) and with relaxed NCF clusters when higher amounts 
of entropy are injected.  This is in good agreement with the constraints placed 
on $S_i$ from analysis of the $L-T$ and $L-M$ relations.  The model also 
qualitatively explains the observed (newly discovered) relationship between 
a cluster's central entropy and the size of its core radius.}

\item{We demonstrate that the model can also successfully explain the 
observed universal temperature profile (for relaxed, massive CF clusters) of 
Allen, Schmidt, \& Fabian (2001).}

\item{At present, there is a dearth of structural information (e.g., entropy 
and temperature profiles) of relaxed NCF clusters from {\it Chandra} and {\it 
XMM-Newton} data.  A large number of the NCF clusters observed with these 
satellites are the sites of ongoing mergers (e.g., Govoni et al. 2004).  It is 
important to note, however, that in most cases {\it these clusters were selected 
because they were known to be mergers} (for the purposes of studying `cold 
fronts', bow shocks, etc.).  However, there are examples of relaxed NCF clusters 
observed previously with {\it ROSAT} and {\it ASCA} (e.g., Buote \& Tsai 1996, 
see also \S 6 of the present study).  A simple way of testing the present model, 
therefore, would be to obtain deprojected entropy profiles for these systems 
(using the current generation of satellites) and see whether or not they possess 
elevated entropy levels at the centers.}    

\end{itemize}

\vskip 0.1in
\noindent The authors wish to thank the referee for useful comments which 
improved the quality of the paper, particularly the discussion of observational 
results.  We thank Alastair Edge, Gilbert Holder, James Binney, and Greg Bryan 
for useful comments on an earlier version of the manuscript.  We also thank 
Steve Allen, Roderick Johnstone, Henric Krawczynski, Ewan O'Sullivan, and Ming 
Sun for providing their Chandra/XMM-Newton results in electronic form and Greg 
Bryan and Mark Voit for providing the entropy profiles of their numerically 
simulated 
clusters.  I. G. M. is supported by a postgraduate scholarship from Natural 
Sciences and Engineering Research Council of Canada (NSERC).  A. B. is supported 
by an NSERC Discovery Grant and M. L. B. is supported by a PPARC fellowship.

\end{document}